# A short review on joint weak and strong cluster lens-mass reconstruction

**Ben D. Normann**[1,a], **Kenny Solevåg-Hoti**[2,b], **Hans Georg Schaathun**[1,2,c]

[1] Department of ICT and Natural Sciences, Norwegian University of Science and Technology, Larsgårdsvegen 2, Ålesund 6009, Møre og Romsdal, Norway

[2] The University Library, Norwegian University of Science and Technology, Larsgårdsvegen 2, Ålesund 6009, Møre og Romsdal, Norway

**Abstract** The distinction between weak and strong lensing is somewhat arbitrary, and both regimes are manifestations of the same physical phenomenon: gravity bending the path of light. Nevertheless, these two regimes have to a large extent been treated separately, since they require different approaches. This review traces the development of methods combining weak-lensing and strong-lensing data for joint lens-mass reconstruction, with a particular emphasis on cluster lenses, where both effects occur. We conclude that so-called inverse methods have been successful in merging the two regimes insofar data analysis is concerned. However, a number of improvements seem to be in place. First, not many studies include weak-lensing data beyond shear. In light of the unprecedented quality of the data of JWST and future surveys, this is a clear point of improvement. Especially so since flexion terms have proven useful in determining substructures. Second, considering the amount of data available, and the complexity of nonparametric lenses, automating the processes of lens-mass reconstruction would be beneficial. Towards this end, invoking machine learning seems like a promising way forward.

## 1 Introduction

The idea of light being bent by gravity goes back at least to Sir Isaac Newton [1] himself, and later John Michell [2] and Pierre-Simon Laplace [3]. But it was not until the twentieth century that it was actually put to use as an observational probe on the universe. Under a solar eclipse in 1905, Eddington and others showed that Einstein's theory of gravity predicted the correct lensing of light, whereas Newton's theory gets it wrong (misses a factor of 2). Today, well over a century later, gravitational lensing is recognized as an important cosmological probe and a rich source of information. One of its uses is the mapping of the (mostly dark) matter in the universe. Mapping the dark matter through gravitational lensing invokes (cluster) lens-mass reconstruction, which will be the focus of this review.

The effect of gravitational lensing is typically divided into two regimes: weak and strong, both of which are important in understanding complicated lensing structures, such as cluster lenses. Since they are both observational aspects of the same phenomenon, it is natural to wonder to what extent they are still treated separately. This paper reviews the literature on different approaches to merging the two regimes into one unified framework for cluster [1] lens-mass reconstruction.

We consider a hybrid approach, combining a structured literature search with a historical approach using prior knowledge and snowball searches. On the one side, we use stringent search criteria, winding up with a total of 32 papers that we examine in some detail. On the other side, we use previous knowledge and snowball searches to complement the historical development. The result is a quasi-historical development.

The rest of the paper is structured as follows. In Sect. 2, we describe our literature search strategy and study selection. In Sect. 3, we go on to describing lensing theory in general, before we focus in on the development of (cluster) lens-mass reconstruction in Sect. 4. This is where the bulk of papers included in the systematic literature review are presented, describing the development up until the state of the art. Finally, in Sect. 5 we present our concluding remarks.

[1] We focus here on cluster lenses, where both effects occur.

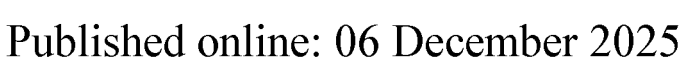

[a] e-mail: ben.d.normann@ntnu.no (corresponding author)
[b] e-mail: kenny.solevag-hoti@ntnu.no
[c] e-mail: hasc@ntnu.no

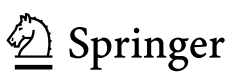

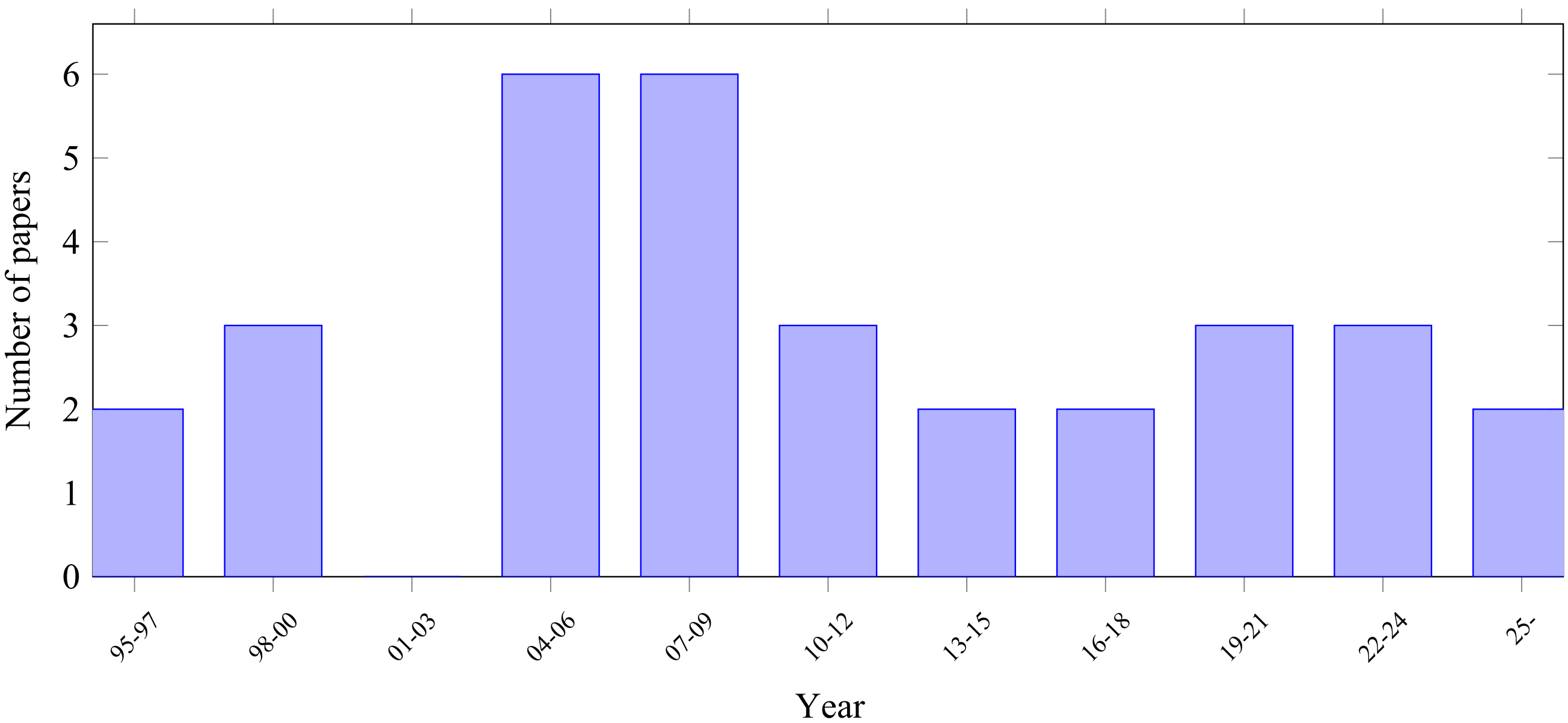

**Fig. 1** The histogram shows the number of papers from the shortlist appearing each year, starting in 1995, and with a bin size of 3 years

## 2 Combining weak and strong: search method

This review considers the joint use of weak- and strong-lensing data for lens-mass reconstruction. To this end, a structured search was conducted, as described below.

### 2.1 Search strategy

To create a search string that would effectively capture results relevant to this review, we went through several iterations. Searches were made in the Web of Science database, chosen for its breadth and reliability. Attention was restricted to *title*, *abstract* and *keywords*, as it was found unlikely that the relevant literature would omit our search words in all these fields. Starting from a broad search, we modified the search string to enforce the right context Fig. 1 and Table 1.

The initial search returned articles from a range of unrelated scientific fields. The next iteration narrowed the hits to our field of interest, but with noise still apparent. The main issue was articles where both weak and strong formalisms were used, but handled independently of each other. Correcting for this, we made use of words present in highly relevant publications in our next search. These hits were relevant, yet limited our scope. Hence, we readjusted for the possibility of authors using a range of synonyms. This lead us to our final query:

(combin* OR unit* OR unify* OR join* OR merg* OR bind* OR link* OR expand* OR extend*) AND weak AND strong AND lens* AND reconstruct*

This search was then applied to the databases Web of Science, Scopus, Compendex and ArXiv[2] providing comprehensive coverage of the field of physics. This resulted in 111 unique papers when cross-checking the results of the different database searches. The breakdown is given in Table 2.

Going through the abstracts, we selected those papers that concentrated on the theoretical framework for weak and strong lensing. Papers that (judging from the abstracts) seemed to be mere observational ones, or papers that concentrated on theoretical treatment of other observables, such as X-rays, were thus discarded. This discarding-by-abstract step was as such a step of calculated risk, as it might sometimes be hard to judge the novelty of the theoretical methodology underlying an observational study. Especially so since a successful theoretical method for observational astrophysics is highly dependent on each individual observational situation. Nevertheless, if a methodology is particularly novel, one would expect it to be presented in a theoretical paper on its own. Consequently, and performed with some care, this step was taken in order to narrow down the scope of the material.

[2] In ArXiv, we refined our search string back to (weak AND strong AND lens* AND reconstruct*) due to the databases search capabilities.

**Table 1** Preliminary searches in the topic field of the Web of Science database 28 October 2022

| Search | String | Hits |
|---|---|---|
| 1 | weak AND strong AND lens* | 1471 |
| 2 | weak AND strong AND lens* AND reconstruct* | 171 |
| 3 | (combin* OR unit*) AND weak AND strong AND lens* AND reconstruct* | 37 |
| 4 | (combin* OR unit* OR unify* OR join* OR merg* OR bind* OR link* OR expand* OR extend*) AND weak AND strong AND lens* AND reconstruct* | 95 |

**Table 2** Number of papers found in each database in the search performed on the 28 October 2022 There are 111 unique papers. In addition to these, 5 papers of interest appeared after this date (in 2023-2025). These have also been included in this review

| Database | Search field | Hits |
|---|---|---|
| Web of Science | Topic | 95 |
| Scopus | Title, abstract, keywords | 67 |
| ArXiv | Abstract | 55 |
| Compendex | Subject, title, abstract | 53 |

### 2.2 Study selection

As a result, a total of 32 papers (two of which appeared after the original search was done) were left for deeper scrutiny. This list contained all the papers that were found in an initiation search,[3] before a structured approach was adopted. The final list of 32 papers is presented in Appendix A. Over the course of the next couple of sections, we discuss all the 32 papers from the list, in a quasi-historical manner. Snowball searches greatly expanding the list have also been adopted, so that the total number of papers referenced is about fivefold that of the shortlist (32) Fig. 2.

It is noteworthy that our systematic search strategy left out a whole strand of relevant papers, with a couple of exceptions. These were the direct methods extending the weak-lensing regime towards the strong regime, by including higher-order effects, such as flexion. Although we have not performed a systematic literature search to include such papers, they are discussed generally in Sect. 3.4 and in the context of cluster lenses in Sect. 4.7.

## 3 Lensing theory

In the following, we present the basic mathematical framework used in state-of-the-art gravitational lensing. The interested reader is referred to standard texts such as the book by Schneider, Ehlers and Falco (often denoted SEF) [4] or Schneider et al. [5] for details, or more recent treatments, like Congdon and Keeton [6]. Finally, consider also the even more recent review by Umetsu, where cluster-galaxy weak lensing is considered [7].

The starting point of gravitational lensing is the so-called ray-trace equation

$$\boldsymbol{\theta} D_{\mathrm{S}} = \boldsymbol{\beta} D_{\mathrm{S}} + \hat{\boldsymbol{\alpha}} D_{\mathrm{LS}}, \tag{1}$$

which relates angular position $\boldsymbol{\theta}$ in the lens plane, at distance $D_{\mathrm{L}}$ from the observer, to corresponding positions $\boldsymbol{\beta}$ in the source plane, at distance $D_{\mathrm{S}}$ from the observer. This equation follows directly from the geometry assumed in Fig. 2, when equipped with an additional assumption of small angles. The geometry typically[4] follows due to a thin-lens approximation, which means that all the mass has been projected down to a two-dimensional plane orthogonal to the optical axis. This is often warranted as the region where the lens affects the light beam is relatively localized, compared to the distances $D_{\mathrm{L}}$ and $D_{\mathrm{S}}$. In fact, most lens-mass reconstruction techniques seek to reconstruct the two-dimensional, lens-plane projected mass density. This reveals a fundamental shortcoming of the current capacity of GL. Albeit being standard procedure, the thin-lens approximation cannot reveal three-dimensional structures of a single lens—at least not in a model-free manner. In reality, lenses are, however, three-dimensional, a structure which could be of importance. Take Morandi et al. [8] as an example, where A1689 is modelled with a triaxial structure. This allows to explain discrepancies between X-ray and strong-lensing data as a result of the internal structure of the cluster. As such, this demonstrates the importance of taking the three-dimensional structure of the lens into account in explaining the data.

Defining the reduced deflection angle by

$$\boldsymbol{\alpha}(\boldsymbol{\theta}) \equiv \frac{D_{\mathrm{LS}}}{D_{\mathrm{S}}} \hat{\boldsymbol{\alpha}}(\boldsymbol{\theta}), \tag{2}$$

[3] Actually, the most interesting papers were already found in our initial search!

[4] Not so with the Schwarzschild lens, where exact results are available.

**Fig. 2** Lensing geometry. $D_L$ and $D_S$ are angular diameter distances from observer to lens and observer to source, whilst $D_{LS}$ is lens to source. $\beta$ and $\theta$ are angular separations of source and image from the optical axis. The light-ray deflection is denoted by $\hat{\alpha}$, whilst $\alpha$ is the reduced deflection angle

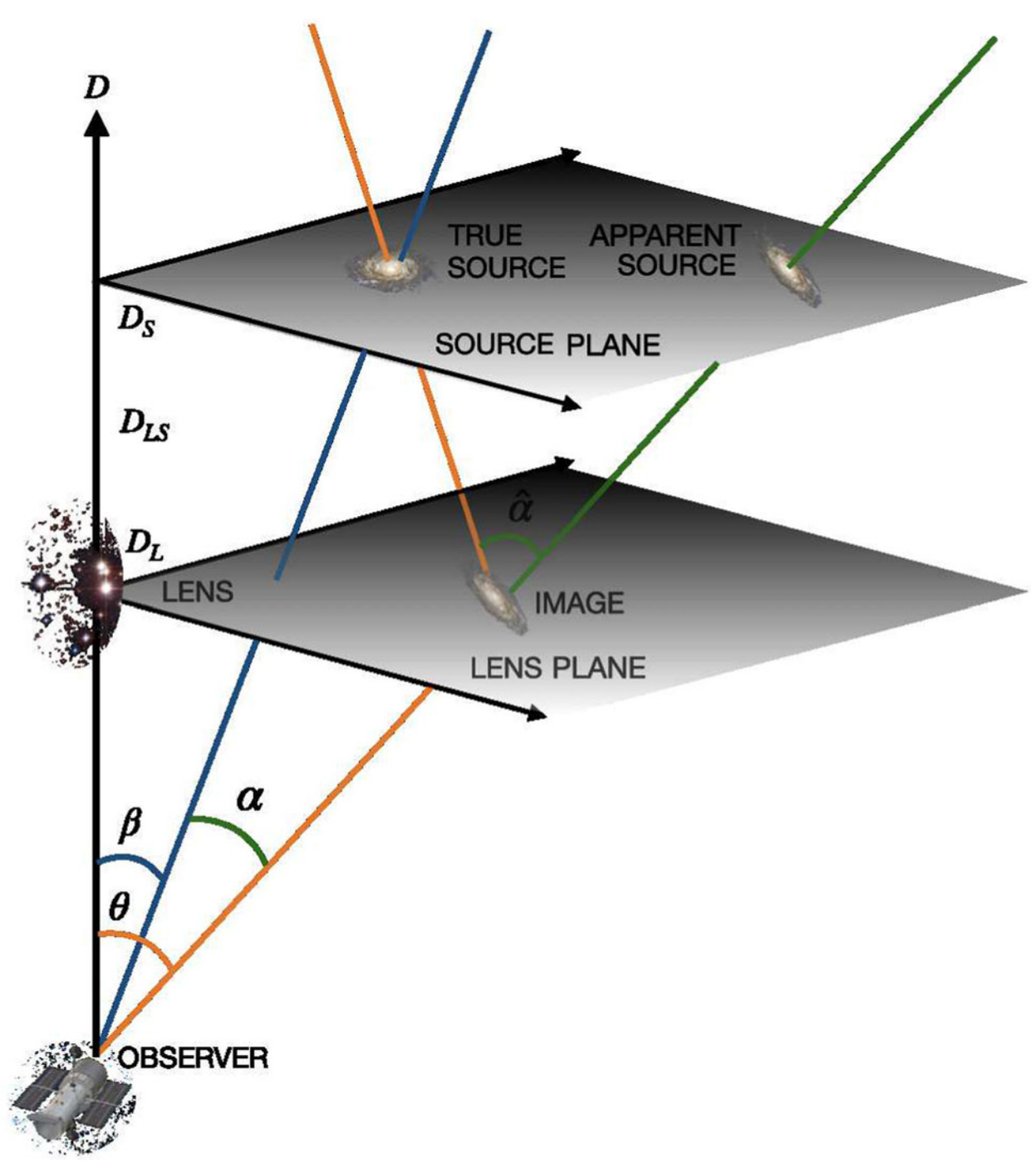

the lens equation is obtained as

$$\boldsymbol{\beta} = \boldsymbol{\theta} - \boldsymbol{\alpha}(\boldsymbol{\theta}). \tag{3}$$

In the thin-lens approximation, one defines the dimensionless surface-mass density $\kappa$ (also called *convergence*) as

$$\kappa(\boldsymbol{\theta}) = \frac{\Sigma(\boldsymbol{\theta})}{\Sigma_{cr}}. \tag{4}$$

Here, $\Sigma$ is the surface-mass density and $\Sigma_{\mathrm{cr}}$ is a critical value defined such that

$$\Sigma_{cr} = \frac{c^2}{4\pi G}\frac{D_S}{D_L D_{LS}}. \tag{5}$$

It is now customary and convenient to define the *lensing potential* $\psi$ (also called the deflection potential) from Poisson's equation for gravity in two dimensions,[5] which takes the form

$$\nabla^2\psi = 2\kappa. \tag{6}$$

One may use Green's functions to show that an integral solution to this equation is given by

$$\psi(\boldsymbol{\theta}) = \frac{1}{\pi}\int_{\mathbb{R}^2} \mathrm{d}^2\theta' \kappa\left(\boldsymbol{\theta}'\right) \ln\left|\boldsymbol{\theta} - \boldsymbol{\theta}'\right|. \tag{7}$$

It follows that the (reduced) deflection angle $\boldsymbol{\alpha}$ may be written as

$$\boldsymbol{\alpha} = \nabla\psi. \tag{8}$$

Using this, the lens equation (3) takes the form

$$\boldsymbol{\beta} = \boldsymbol{\theta} - \nabla\psi(\boldsymbol{\theta}). \tag{9}$$

Consequently, for a given lens $\psi$, Eq (9) is a mapping $\mathcal{L}_\psi : \mathcal{U} \rightarrow \mathcal{D}$, where $\mathcal{U}$ and $\mathcal{D}$ are sets of undistorted and distorted images, respectively. Since distorted background sources are typically much smaller than the angular scale on which the lens properties change, the mapping can be locally linearized and the lens mapping (9) may be written as

$$\boldsymbol{d\beta} = \mathcal{A}\,\boldsymbol{d\theta} \tag{10}$$

where the so-called *amplification matrix* $\mathcal{A}$ is the Jacobian of the coordinate transformation. By way of (9), we have

$$\mathcal{A}(\boldsymbol{\theta}) = \frac{\partial\boldsymbol{\beta}}{\partial\boldsymbol{\theta}} = \left(\delta_{ij} - \psi_{ij}\right) \quad \text{where} \quad \psi_{ij} \equiv \frac{\partial^2\psi}{\partial\theta^i\partial\theta^j}. \tag{11}$$

[5] The lens plane.

Any Hermitian $2 \times 2$ matrix can be decomposed into the identity matrix and the Pauli spin matrices:

$$\mathcal{I} = \begin{pmatrix} 1 & 0 \\ 0 & 1 \end{pmatrix} \quad , \quad \sigma_1 = \begin{pmatrix} 0 & 1 \\ 1 & 0 \end{pmatrix} \quad , \tag{12}$$

$$\sigma_2 = \begin{pmatrix} 0 & -i \\ i & 0 \end{pmatrix} \quad , \quad \sigma_3 = \begin{pmatrix} 1 & 0 \\ 0 & -1 \end{pmatrix}. \tag{13}$$

It follows from (9) that $\mathcal{A}$ takes the form

$$\mathcal{A} = (1-\kappa)\mathcal{I} - \gamma_\times \sigma_1 + \mathrm{i}\rho\sigma_2 - \gamma_- \sigma_3, \tag{14}$$

where standard theory yields $\rho = 0$, since gravity is curl-free at the level of approximation used (refer to Bacon and Schafer [9] for an interesting discussion). In the above, we have defined the (complex) *shear*

$$\gamma = \gamma_+ + i\gamma_\times \tag{15}$$

with (real) components

$$\gamma_+ = \frac{1}{2}(\psi_{11} - \psi_{22}), \quad \gamma_\times = \psi_{12} = \psi_{21}. \tag{16}$$

By defining the *reduced shear*,

$$g \equiv g_+ + i g_\times \tag{17}$$

with components

$$g_+ \equiv \frac{\gamma_+}{1-\kappa} \quad , \quad g_\times \equiv \frac{\gamma_\times}{1-\kappa}, \tag{18}$$

one could alternatively express the amplification matrix as

$$\mathcal{A} = (1-\kappa)\begin{pmatrix} 1-g_+ & -g_\times \\ -g_\times & 1+g_+ \end{pmatrix}. \tag{19}$$

From the latter form of $\mathcal{A}$, it is particularly apparent that $(1-\kappa)$ is a scaling factor, whereas the (reduced) shear alters the shape of the image (by elongation). Consequently, based on the assumption of randomly oriented intrinsic ellipticities in the sources ,[6] measurements of image ellipticities yield an estimate of the reduced shear.

### 3.1 Direct lens-mass reconstruction

The shear may be used to obtain a measure of the surface-mass density, as follows. Equations (7), (15) and (16) together give

$$\gamma(\boldsymbol{\theta}) = \frac{1}{\pi}\int_{\mathbf{R}^2} \mathrm{d}^2\theta' \mathcal{D}\big(\boldsymbol{\theta} - \boldsymbol{\theta}'\big)\kappa\big(\boldsymbol{\theta}'\big), \tag{20}$$

with

$$\mathcal{D}(\boldsymbol{\theta}) \equiv \frac{\theta_2^2 - \theta_1^2 - 2\mathrm{i}\theta_1\theta_2}{|\boldsymbol{\theta}|^4}, \tag{21}$$

That is, $\gamma$ is a convolution of $\kappa$ with kernel $\mathcal{D}$, a relation which may be inverted to give

$$\kappa(\boldsymbol{\theta}) - \kappa_0 = \frac{1}{\pi}\int_{\mathbf{R}^2} \mathrm{d}^2\theta' \mathcal{R}\mathrm{e}\big[\mathcal{D}^*\big(\boldsymbol{\theta} - \boldsymbol{\theta}'\big)\gamma\big(\boldsymbol{\theta}'\big)\big]. \tag{22}$$

This relation opens up the possibility of obtaining matter maps of the universe by *directly* measuring the shear of distantly incoming light congruences. However, shear in the weak-lensing regime is a relatively weak distortion, and thus relatively hard to detect. In fact, it is necessary to average over large samples of galaxies for the effect to appear. Obtaining the surface-mass density by way of shear measurements has been extensively applied since its first measurement by Tyson et al. [10]. A more modern example is given by Jauzac et al. [11].

[6] Or otherwise a known distribution of ellipticities.

*3.1.1 Kaiser–Squires inversion*

Obtaining the lens mass via other observables, such as the shear, is known as *lens-mass reconstruction*. The relation (22) was used by Kaiser and Squires [12] in developing a parameter-free[7] inversion technique to obtain the surface-mass density from shear measurements. This inversion technique—hereafter simply referred to as the *KS inversion technique*—quantifies the shear using the quadrupole-moment tensor of the surface brightness(cf. Miralda-Escude [13]; Miralda-Escude [14]). Following Congdon and Keeton [6, Chapt. 7], the quadrupole-moment tensor is given by

$$Q_{ij} = \frac{\int_{\mathbb{R}^2} (\theta_i - \bar{\theta}_i)(\theta_j - \bar{\theta}_j) I(\boldsymbol{\theta}) \mathrm{d}^2\theta}{\int_{\mathbb{R}^2} I(\boldsymbol{\theta}) \mathrm{d}^2\theta}, \tag{23}$$

where $\bar{\boldsymbol{\theta}} = (\bar{\theta_1}, \bar{\theta_2})$ is the image centroid position, given by

$$\bar{\boldsymbol{\theta}} = \frac{\int_{\mathbb{R}^2} \boldsymbol{\theta} I(\boldsymbol{\theta}) \mathrm{d}^2\theta}{\int_{\mathbb{R}^2} I(\boldsymbol{\theta}) \mathrm{d}^2\theta}. \tag{24}$$

Note that the KS inversion technique hinges on the assumption that the overall orientation $\Xi$ of each galaxy is randomly distributed[8] (ensemble average), so that there is no intrinsic alignment; that is,

$$\langle \Xi^2 \rangle = 0. \tag{25}$$

The validity of such an assumption is of course questionable, as reviewed in Troxel and Ishak [15]. Making use of the assumption (25), one may, however, compare source and image ellipticities, by such providing a measure of the (reduced) shear. This is in turn inverted to obtain the surface-mass density of the lens. A problem with the method is that it is valid only for sufficiently weak lensing, where the (complex) distortions $\delta$ produced by the lens are linearly related to the shear. In particular, one has

$$\delta \approx 2g \approx 2\gamma \qquad \text{linear regime.} \tag{26}$$

The method is later extended to the nonlinear regime by Schneider and Seitz [16, 17], relieving the assumption of weak lensing, as is necessary to probe, for example, the centre of clusters. The authors show that in the generalized formulation, the shear is not an observable quantity. The only local observable from image distortions is the (complex) distortion $\delta$, given by

$$\delta = \frac{2g}{1 + |g^2|}, \qquad \text{nonlinear regime} \tag{27}$$

with $g$ given in Eq. (17). Evidently, Eq (27) reduces to (26) for $\kappa$, $\gamma \ll 1$. Another problem with the KS inversion technique is that the inversion formulae is exact only if one assumes observational data on the whole lens plane. The finitude of the CCD or the data field thus causes problems, an issue discussed in various ways by several authors [18–21]. Finally, redshift information is taken into account Seitz and Schneider [22]. An improved finite-field inversion method similar to the one developed in the aforementioned literature is also presented in [23]. Therefore, in summary, the KS method had before the end of the millennium thus been improved to account for

a. Strong tidal fields (nonlinear regime) in cluster centres.
b. Finite data fields.
c. Redshift distribution of background galaxies.

The resulting methods were computationally relatively efficient. Also, they were local, in the sense that the surface-mass density is obtained from observed ellipticities of background galaxies. However, as discussed by Seitz et al. [24], a number of drawbacks were also present:

- Whilst the data need to be smoothed, objective criteria for how to set the smoothing scale are lacking.
- The quality of the reconstruction is hard to quantify.
- Constraints from additional observables, such as multiple images or arclets[9], are not easily included.
- The magnification, which is needed to break the so-called mass-sheet degeneracy (to be discussed), is incorporated in a global fashion.

Finally, note also the so-called quadruple method introduced by Lombardi and Bertin [25], where the isotropy requirement (25) is replaced by an isotropy requirement on the observable quadrupole moments instead. Note also the works on improving the accuracy of mass reconstruction in various ways (Lombardi and Bertin [26], Lombardi et al. [27]). Also, mass reconstruction is considered through a variational formulation in Lombardi and Bertin [28], where the procedure formulated is reported to reduce the relevant execution time by a factor of 100–1000 with respect to the fastest methods available at the time.

[7] Or nonparametric. This simply means that no a-priori model for the lens is made, as further discussed in a later section.

[8] As a side remark: Let it be mentioned that also Tyson et al (1990) assumed that the net alignment of the source galaxies was zero in their detection prior to the work by Kaiser and Squires.

[9] *Arclet* seems to be the term used for sheared (stretched) images, whereas the term *arc* is reserved for stronger (flexed) distortions.

### 3.2 Inverse methods for lens-mass reconstruction

The aforementioned issues with the standard KS method, or direct methods more broadly, lead to the development of so-called inverse methods, such as maximum-likelihood methods (Squires and Kaiser [29], Bartelmann et al. [30]). In the inverse methods, one seeks to fit a very general lens model with data. One typically parametrizes the lens potential on a grid and then minimizes the regularized log-likelihood function. Take $N_{\mathrm{g}}$ to be the number of galaxies and $\epsilon_i$ to denote ellipticity of the $i$-th galaxy. The shear-likelihood function is then given by Schneider et al. [5, Part 3, Eq. 62]

$$-\ln \mathcal{L} = \sum_{i=1}^{N_{\mathrm{g}}} \frac{|\epsilon_i - g(\boldsymbol{\theta}_i, \psi_n)|^2}{\sigma_i^2(\boldsymbol{\theta}_i, \psi_n)} + 2\ln\left(\sigma_i(\boldsymbol{\theta}_i, \psi_n)\right) + \lambda_{\mathrm{e}} S(\psi_n), \tag{28}$$

where $\sigma_i$ are standard deviations in the ellipticities, $\boldsymbol{\theta}$ are angular positions, $g$ is the reduced shear, and $\psi$ is the discretized potential. The parameter $\lambda_{\mathrm{e}}$ is a Lagrangian multiplier giving the relative weight of the regularization and the likelihood function. In particular, note that strong-lensing constraints may now be added as Lagrangian multipliers. Thus, in such a formulation of the problem, other types of constraints (e.g. strong lensing and magnification) are readily included in a linear fashion. As an early example, consider the work by Bartelmann et al. [30] in which (inverse) magnification constraints ($r_i$) are included (to effectively break the mass-sheet degeneracy). The $\chi^2$-function

$$\chi^2 = \sum_{k,l} \left( \frac{(g_i(k,l) - \hat{g}_i(k,l))^2}{\sigma_{\mathrm{g}}(k,l)} + \frac{(r_i(k,l) - \hat{r}_i(k,l))^2}{\sigma_{\mathrm{r}}(k,l)} \right) \tag{29}$$

is minimized numerically using a conjugate gradient method. As before, $g_i$ refers to the reduced shear, $\sigma_p$ denotes variance in data of type $p$, and hats ( $\hat{}$ ) are used to signify estimators.[10] The method allows for straightforward inclusion of measurement inaccuracies and is designed to work well with clusters, where the observed field is small or irregularly shaped. This method was extended by Seitz et al. [24] to include regularization of the potential and use the individual galaxy ellipticities and positions instead of averaging by dividing the image into cells. Bridle et al. [31] also include magnification constraints by a maximum-likelihood approach, here seen as a special case of the maximum-entropy method (MEM) in the context of Bayes' theorem. This method allows for direct reconstruction of the mass distribution instead of going via the gravitational potential.

Other log-likelihood approaches include that of Marshall [32], in which an atomic-inference procedure is adopted. Only WL effects are considered, but incorporation of SL effects is reported to be straightforward. Beyond that, the reader is referred to the book-sized review by Bartelmann and Schneider [33] on the early developments of weak gravitational lensing, and weak-lensing theory in general.

As we shall see in Sect. 4, inverse methods have become the standard tool for lens-mass reconstruction. The key factor for its success is its linear problem formulation, where both weak and strong constraints (and others) can be added as linear constraints in a joint analysis.

#### *3.2.1 Degeneracies in the lens equation*

It was early recognized that the lensing equation suffers from a number of degeneracies, owing to the fact that most observables of gravitational lensing are dimensionless. Early sources on this include Falco et al. [34] and Gorenstein et al. [35], both written in a time when cluster lenses were still controversial. Later, papers, such as the one by Saha [36], rederive the degeneracies, putting them in a new observational context. Herein, so-called similarity degeneracies are broadly categorized as either (i) *distance degeneracy* or (ii) *angular degeneracy*. Another type, the so-called *mass-sheet degeneracy*, also appears for sources at the same redshift.[11] In this case, the problem is that the observable distortion, $g = \gamma/(1-\kappa)$, is invariant under the transformation

$$\kappa \to (1-\lambda) + \lambda\kappa, \quad \gamma \to \lambda\gamma. \tag{30}$$

Globally, this is equivalent to the rescaling

$$\mathcal{A} \to \lambda\mathcal{A} \quad , \quad \lambda = \text{const.} \neq 0, \tag{31}$$

which leaves the critical curves $\det \mathcal{A} = 0$ invariant. (Refer to treatments like those of Saha and Williams [38], Liesenborgs and De Rijcke [39] for more.) This means that several source positions may give rise to the same observed position and give and take a mass sheet of constant density. In particular, this transformation keeps the critical curves of the lens mapping fixed, and thus the location of the giant arcs and arclets. The mass-sheet degeneracy is an example of a broader class of source-plane transformation issues for strong lenses, as discussed by Schneider and Sluse [40]. For an even more inclusive discussion of degeneracies, consider also the one by Wagner [41]. Consequently, since one cannot distinguish between lenses when using distortion data alone (for sources at the same redshift), lens-mass reconstruction algorithms must therefore take measures to break the degeneracy through some means or another. One way to go about it, would be to use magnification, as tried by Broadhurst et al. [42], Broadhurst et al. [43]. However, Schneider et al. [44] point out uncertainties in unlensed source counts, leaving the magnification method less prosperous. Bradač et al. [45] set out to use redshifts instead and conclude that it is necessary to simultaneously analyse weak and strong lensing to break the mass-sheet degeneracy. Thus, it is about time that we give a more formal definition of the difference between the weak- and strong-lensing regimes.

[10] Note that the originally published manuscript seems to have a misprint here, using boldface letters instead of hats seemingly.

[11] Note also the generalization discussed in [37].

### 3.3 The difference between weak and strong lensing

Strong lensing refers to conditions allowing for multiple images of the same source (Congdon and Keeton [6, Chapter 2]). For this to happen, the lens must possess a certain 'strength'. The divide between weak and strong lensing is arguably somewhat arbitrary, since it is all part of the same underlying physical phenomena. Nevertheless, from the mathematical side one may show that a sufficient[12] requirement for producing several images of a source under the mapping (1) is $\kappa(0) > 1$, a condition known as *supercriticality*. Furthermore, it may be shown that the magnification diverges whenever $\det \mathcal{A} = 0$, producing the characteristic arcs associated with GL. We can infer the following.

- *Strong-lensing data* ($\kappa > 1$): Incorporating multiple imaging and/or critical curves ($\det \mathcal{A} = 0$).
- *Weak-lensing data* ($\kappa < 1$): This means incorporating shear measurements in the analysis, and sometimes flexion (to be discussed). Note that the aforementioned linear regime ($\kappa, \gamma \ll 1$) lies in the weak-lensing regime.

The reason why these effects have been studied separately is to some extent historical, owing to the methods and sensitivity available. On the one hand, there is the rare but visually apparent effect of strong lensing. This effect is visual and observed as giant luminous arcs or multiple images. The sensitivity required of the instruments to capture this effect is therefore low compared to that of weak lensing, which must typically be treated statistically over large samples of lensed images. Whereas strong lensing was used for lens-mass estimation already with Zwicky [46, 47], weak lensing thus had its dawn in the 90 s, as described earlier.

### 3.4 Beyond shear: Banana or jelly bean?

The lensing theory described in the beginning of this chapter hinges on the map given in Eq. (9), with the linearization given by (10). For infinitesimal beams, a linearization is warranted, but if one intends to take into account the finitude of the beam, the linearization breaks down. Put differently, if the shear varies over the cross section of the image, higher-order effects, such as flexion and second-flexion, come into play. First-flexion is a shift in position of the image centroid, whereas second-flexion creates the banana-looking shape.[13] The flexion is a second-order[14] effect, and thus weaker than shear in the weak field. On the one hand, this could cause problems in detecting it. On the other hand, however, source galaxies are not believed to be intrinsically flexed, which provides an advantage: the flexion generally belongs to the image. One way to quantify the flexion of a lensed image is by multipole moments, as, for instance, in the early work by Goldberg and Natarajan [48], in which it is shown how flexion (octopole moment) can be used as a probe of weak-lensing shear fields. Note also later contributions such as those by Goldberg and Leonard [49], Okura et al. [50], Okura et al. [51]. Another way to quantify flexion is through the shapelets formalism due to Refregier. This was used by Goldberg and Bacon [52], where an inversion technique based on flexion was investigated, reporting an increased signal-to-noise ratio. The flexion formalism is tidily presented in complex form by Bacon et al. [53], also revealing a spin-3 field which has come to be known as the aforementioned second-flexion. To summarize, one has

$$\mathcal{F} = |\mathcal{F}|\mathrm{e}^{\mathrm{i}\phi} = \partial\kappa = \partial^*\gamma \qquad \text{first flexion (spin-1)}, \tag{32}$$

$$\mathcal{G} = |\mathcal{G}|\mathrm{e}^{\mathrm{i}3\phi} = \partial\gamma \qquad \text{second flexion (spin-3)}, \tag{33}$$

where $\partial = \partial_1 + i\partial_2$ is a complex derivative incorporating the two directions (1, 2) of the screen-space and star (*) denotes complex conjugation. Sometimes $\mathcal{F}$ is referred to as comatic flexion, and $\mathcal{G}$ is referred to as trefoil flexion. The flexion has four components, and the relation between the spin-1 and spin-3 components is given by

$$\partial^*\partial\mathcal{G} = \partial\partial\mathcal{F} \qquad \text{consistency relation}, \tag{34}$$

which can be readily proven. The paper also calculates the flexion coefficients for various spherical lensing profiles. Later, in Bacon and Schafer [9], it is also shown how the flexion term arises when one includes the final Pauli spin matrix.[15] The reader should also note the seemingly independent development of higher-order lensing throughout the papers by Irwin and Schmakova [54–57], under the name *sextupole* lensing.[16] Although originally intended for galaxy–galaxy lensing, flexion measurements were already

[12] Sufficient and necessary in the case of axisymmetric lenses. Only sufficient in the more general case.

[13] or is it a jelly bean?

[14] If we by 'order' mean the number of derivatives in a Taylor-expansion.

[15] That is; nonzero $\rho$ in Eq. (14)

[16] In contrast to Goldberg's notion of flexion as an octopole term.

in 2007 performed on the cluster Abell 1689, revealing a dark-matter structure that was not apparent from shear measurements alone (Leonard et al. [58]). Flexion can also be estimated through the so-called analytic image model (AIM, Cain et al. [59]), and a subsequent study based on this method confirms that significant substructures may remain invisible if this effect is not taken into account in cluster lensing, cf. Cain et al. [60]. Flexion with elliptical lens profiles was investigated by Hawken and Bridle [61], where it is found that the constraints on galaxy halo ellipticities are comparable or even up to two orders of magnitude higher than those from shear.

Just like the actual observable of shear $\gamma$ is the reduced shear $g$, the actual observable when it comes to flexion is the *reduced flexion*, calculated and tested against simulations by Schneider and Er [62]. One result is that the presence of flexion affects the determination of the reduced shear. Er et al. [63] consider the problem of cluster lens-mass reconstruction, combining strong-lensing constraints with weak-lensing shear and flexion. By so, the weak-lensing analysis is extended to the inner parts of the cluster, and the resolution of substructure is improved. Also, Lasky and Fluke [64] calculate the flexion for a diversity of circularly symmetric lenses and conclude that the convergence and first-flexion are better indicators of the Sérsic shape parameter, whilst for the concentration of NFW profiles, the shear and second-flexion terms are preferred. In a follow-up work by Fluke and Lasky [65], the ray-bundle method (RBM) developed by Fluke et al. [66] is applied and found to accurately recover second-flexion for the Schwarzschild lens. Also, the existence of a preferred flexion zone is demonstrated. The RBM could be described as an attempt at using a strong-lensing approach to weak lensing, by considering the fact that light rays come in bundles and not as single, infinitesimal rays. The idea of a strong-lensing formalism for weak lensing is furthered by Fleury et al. [67–69]. Here, the *infinitesimal-beam approximation* underlying weak-lensing approaches is found to yield problems. This approximation is shown to comprise the following three approximations:

- *Flat-sky-approximation*: Distances are such that the lens and source planes can be approximated by planes.
- *Weak-field regime*: The space-time metric can be considered locally flat within the beam's cross section.
- *Smooth curvature*: The Riemann curvature can be considered constant over the beam's cross section.

By rejecting this approximation and studying instead a finite beam, the authors show that the Kaiser–Squires theorem is violated by finite-size effects.

In another paper, this time by Birrer et al. [70], the two regimes (weak and strong) are sought united through *line-of-sight* (LOS) deflectors in galaxy-scale lensing. In this formalism, the main deflector is a strong deflector, upon which small LOS deflectors are added in a perturbative manner. Strong-lens systems are shown to be accurate and precise probes of cosmic shear.

The converse, weak-lensing approach to strong lensing is taken by Clarkson [71], where the general theory of secondary weak gravitational lensing is developed. This approach is later extended to include all leading-order screen-space derivatives (Clarkson [72, 73]), as such extending the work by Bacon et al (2006 and 2009) from an altogether different route. This latter extension by Clarkson is referred to as the *roulette formalism*, and for circularly symmetric lenses, it is shown to reproduce strong-lensing features. Recursion relations (consistency relations) like Eq. (34) are later developed by Normann and Clarkson [74] for the higher-order terms in context of general nonsymmetrical lenses. These relations proved useful in a simulator (CosmoAI) developed by Schaathun et al. [75], in which a visualization feature for show-casing the roulette expansion may be found. This also serves as a probe to investigate the simulators ability for lens-mass reconstruction, as described by Schaathun et al. [76]. It is noteworthy that the roulette formalism in its current standing does not include all terms in the full, nonlinear geodesic deviation equation, as described by Vines [77].

Finally, note also the work by Lanusse et al. [78], where a mass-mapping algorithm designed to recover small-scale information from a combination of gravitational shear and flexion is developed, starting from the geodesic deviation equation. The reconstruction method is tested on realistic weak-lensing simulations corresponding to typical Hubble Space Telescope (HST)/Advanced Camera for Studies (ACS) cluster observations. The inclusion of flexion is found to increase signal-to-noise level on small scales.

It seems fair to conclude that despite difficulties in actually measuring flexion (cf. Viola et al. [79], Rowe et al. [80]), this higher-order effect has proven to be of importance. Although perhaps limited by the current state-of-the-art observational situation, it is natural to wonder if even higher-order terms in the roulette formalism could prove useful.

### 3.5 Cluster lenses

The observation of weak lensing by clusters was reported by Tyson, Valdes and Wenk [10] in a seminal paper, in some sense marking the beginning of observational support for weak lensing. Attempts to determine cluster lens-mass distributions, however, seem to go back to Webster [81], whilst Kochanek [82] and Miralda-Escude [14] discussed how parameterized cluster-mass distributions could be constrained from weak-lensing data through ellipticities of the source and image. The effects of cluster strong lensing were, however, first unknowingly discovered by Lynds and Petrosian [83] when observing giant luminous arcs,[17] with subsequent interpretation of the phenomena as a lensing effect by Soucail et al. [84].

[17] Zwicky also considered clusters, but at this stage in the development shift in position was the discussed effect.

Whereas a power law of the form $\rho \sim 1/r^{\eta}$ seems to work well for modelling lenses on galactic scale, this does not apply to clusters. A natural generalization would be to allow the power law index to take on one value for small radii and another for large radii, like the Navarro–Frenk–White (NFW) model. Here,

$$\rho(r) = \frac{\rho_s r_s^q}{r^p (r_s + r)^{q-p}} \qquad \text{NFW profile,} \tag{35}$$

for some scale radius $r_s$. For $r \ll r_s$, one finds $\rho \sim 1/r^p$, whereas $\rho \sim 1/r^q$ for $r \gg r_s$. From cosmological simulations of cluster-sized dark-matter halos, Navarro, Frenk and White [85–87] found $p = 1$ and $q = 3$. Later parametric models have often taken this profile as a starting point. Now that the resolution of the image data is increasing, there is, however, a need for going beyond such parametric models.

### 3.6 Parametric and nonparametric (free-form) models

For the purpose of this review, it is useful to make a distinction between parametric and nonparametric models. We employ the classification from Lefor et al. [88], where parametric models make assumptions on physical objects (e.g. point-masses and singular isothermal spheres) to model the lensing mass. Whilst nonparametric (or free-form) models do lens inversion by avoiding such assumptions and rely on the lensed images alone to do mass reconstruction. In the nonparametric approach, several models can fit the observed data, and a regularization problem needs to be solved to restrict results to plausible surface-mass densities. An example of a nonparametric approach that created some stir is the one by Jee et al. [89], where a dark-matter ring in the cluster CL0024+17 ($z = 0.395$) was discovered, using both strong- and weak-lensing constraints. When using nonparametric models, a large number of constraints are needed to avoid spurious results, as discussed, for instance, by Ponente and Diego [90] and Sendra et al. [91].

Abell1689 is a good example of a cluster where nonparametric models are competitive with parametric ones, since there are so many constraints (hundreds of arcs). Parametric models are typically popular in the inner parts of clusters, where the lensing effects are strong. In practice, a combination of both parametric and free-form modelling is used (cf., e.g. Jullo and Kneib [92] and Beauchesne et al. [93]).

### 3.7 The point-spread function (PSF)

Although not a focus in this review, one might also consider the crucial importance of the point-spread function (PSF) in weak-lensing studies, as discussed by Abriola [94]. The PSF models how the image of a point-source is affected by the optical system (telescope) and the atmosphere. In order to analyse the shear field of a lensing cluster, one must know the shape of background galaxies. Since stars in the background field are point-like sources, one may use these to assess the quality of the PSF: if these stars appear elongated, this is likely to be due to the PSF. The mentioned study analyses the cluster Abell 2744 with weak lensing. It also compares with SL results from the literature, reporting great consistency. However, no joint analysis of the data seems to have been performed. Joint analysis is, however, the focus of the next section.

## 4 Cluster lens-mass reconstruction from weak and strong lensing combined

The treatment so far has been general, giving the development of weak-lensing theory beyond shear (i.e. towards the strong regime) and by explaining important concepts and issues. Furthermore, the turn to maximum-likelihood methods (discussed in Sect. 3.2) was an important step which the community continued to develop from Bridle (1998, Sect. 3.2) and on. With it, both weak and strong constraints could be jointly analysed through linear constraints.

In the current section, we narrow in to cluster lens-mass reconstruction and review the development of methods for unified analysis. The shortlist from the systematic review (cf. discussion in Sect. 2) makes up the bulk of material covered.

### 4.1 Weak lensing (arclets) combined with multiple images

The strong-lensing regime permits for multiple images of the same source. Such information may be used for lens-mass reconstruction, as investigated by numerous sources, such as AbdelSalam et al. [95]. This particular work by Abdel Salam et al., however, is furthered in a series of papers to include also weak-lensing information for cluster lens-mass reconstruction (AbdelSalam et al. [96, 97], Saha et al. [98]). The technique developed uses constraints from multiple images and arclets. The technique bears resemblance with that of Kaiser and Squires [12] for weak lensing, but this time the problem formulation is *linear*: the aforementioned strong-lensing constraints are included as linear constraints on the projected mass distribution, together with weak-lensing constraints. Quadratic programming is then used to obtain the mass maps, and the method developed is reported to overcome the drawbacks of nonlinear methods, such as the mass-sheet degeneracy (by using sources at different redshifts), and the technique is also applied to cluster lens-mass reconstruction (see also AbdelSalam et al. [99]).

Another early example is Kneib et al. [100], which studies the particular cluster C1 0024+1654 ($z = 0.395$). The study incorporates strong-lensing constraints (multiple images) in a weak-lensing analysis, using the maximum-entropy method developed by Bridle (1998, Sect. 3.2). The strong-lensing constraints are found to narrow down the parameter space effectively, with an analysis strongly rejecting the SIS model, whilst fitting that of an NFW profile well. A shortcoming of the study is that it is not parameter-free. Also, the parametrization only allows weak-lensing signals in the outer parts ($\sim$ 570 kpc) and strong-lensing constraints from the inner regions (155 kpc). In Smith et al. [101], the software LENSTOOL is used together with additional routines to incorporate weak-lensing constraints. Again the algorithm used builds on parametric models, with a $\chi^2$-estimator quantifying how well each trial lens model fits the data.

With these works in mind, it therefore seems fair to conclude that around the dawn of the twenty-first century, the theory was sufficiently mature and observations crisp enough for a joint analysis of both weak- and strong-lensing constraints. For it is one thing to master joint analysis in theory, and quite another to implement it with actual observations. As that step was overcome, however, combining the two regimes is quickly confirmed to be beneficial. But, a number of theoretical problems remained, however, in need of closer attention:

(i) The mass-sheet degeneracy must be broken for any study relying on the lens equation,
(ii) Likelihood methods must be properly regularized.
(iii) Nonparametric models should be developed.

### 4.2 Bradač, Umetsu and mass-sheet degeneracy

As already discussed (cf. Sect. 3.2.1), attempts had been made to overcome the mass-sheet degeneracy, which also became an issue in terms of discriminating between the SIS and the NFW profile—a hot topic in connection with cluster lenses. In 2004, Bradač et al published a paper on the mass-sheet degeneracy [45], suggesting a way to break it by using the distortion and the redshift information of background galaxies. They conclude that the method is effective for critical clusters only (multiple imaging). Consequently, in order to break the degeneracy with the current (2004) data, it was found necessary to extend the statistical lensing analysis closer to the cluster centre, performing a joint weak- and strong-lensing analysis. An overview of the results is presented in a proceedings [102], with reference to further literature, alongside an implementation with the cluster-mass distribution of RX J1347-1145. The method developed is again parametric, and attention is called to nonparametric models, since the method in principle should be applicable in such scenarios as well. Consequently, in a series of papers, Bradač et al propose a nonparametric method for uniting weak and strong lensing [103, 104], accompanied by observational implementations [105–107]. The method relies on minimizing the $\chi^2$ functional

$$\chi^2(\psi_k) = \chi^2_{\mathrm{WL}}(\psi_k) + \chi^2_{\mathrm{MI}}(\psi_k) + \eta R(\psi_k), \tag{36}$$

where subscripts WL and MI denote weak lensing and multiple imaging, respectively. $\psi_k$ is the lensing potential on a regular grid, whereas the final term is a regularization term. A minimum is sought such that

$$\partial \chi^2(\psi_i)/\partial \psi_i = 0. \tag{37}$$

The authors comment on the fact that other attempts at combining weak and strong lensing have already been made. For instance, they point out that Abdelsalam et al and Smith et al (Sect. 4.1) use the surface-mass density $\kappa$, on which the shear $\gamma$ and the deflection angle $\alpha$ depend nonlocally. Bradač et al use instead the potential $\psi$, from which both convergence, shear and the deflection angle may be locally derived (derivatives). Where Kneib et al (2003) and Smith et al (2004) (cf. Sect. 4.1) used a parametrized, Bayesian approach, Bradač et al apply a more general parametrization. In 2009, they apply their techniques on the Bullet cluster, with a strong- and weak-lensing analysis following the algorithm first proposed by Bartelmann et al (1996) and discussed in Sect. 3.2.

Another interesting strand of work, based on the authors original idea of using magnification to break the mass-sheet degeneracy (Broadhurst et al. [42]), is also further developed (cf., e.g. Umetsu and Broadhurst [108], Medezinski et al. [109], Umetsu et al. [110]). Finally, a model-free reconstruction method is presented by Umetsu [111], using multiple populations of background sources, correcting for magnification bias and incorporating strong-lensing constraints. The method is applied to a number of clusters from the Subaru images.

### 4.3 Perturbative approach

Another interesting approach within the framework of maximum-likelihood approaches is the perturbative approach undertaken by Natarajan et al. [112]. Again weak and strong are combined as linear constraints in a maximum-likelihood manner. Galaxy–galaxy lensing in the cluster CL 0024+16 at $z = 0.39$ is used to track the fate of dark-matter subhalos. The gravitational potential is modelled such that

$$\phi_{\mathrm{tot}} = \sum_n \phi_s + \sum_i \phi_{p_i}, \tag{38}$$

where the two $\phi_s$ ($n = 1$ and $n = 2$) components represent smooth large-scale components, whereas the galaxy subhalo components $\phi_{p_i}$ are treated as perturbers. The reason for having two large-scale components is that earlier studies reveal this to be the best-fit model for that particular cluster. Thus, this is taken as a prior in the current study. The results provide strong support for the tidal-stripping hypothesis.

### 4.4 Diego et al and the regularization problem

Due to scarce observations, parametric models were initially popular, and superior to nonparametric ones, where the freedom in the models must be regularized—a process that either requires parameters or a large number of constraints. As observations increased in amount and accuracy, however, constraints from weak and strong lensing combined made such models competitive with parametric ones. This seems to have accelerated the combinations of the two lensing regimes (WL and SL). In addition to the work by Bradač (previous section) and also that by Broadhurst ([42] discussed in Sect. 3.2), the work of combining the two regimes was furthered by Diego et al. [113] in a paper known as Paper III in a series. The work centres around the idea of expanding the projected cluster-mass distribution into a set of basis functions which are then constrained individually using shear and SL data. As a note to the side, the method was also extended by Sendra et al. [91] to take member-galaxy light distributions into account, which surprisingly diminished the need for regularization (at least for SL data).

Again, a maximum-likelihood approach is taken, well suited for the task of combining different types of constraints. Their discretized problem formulation is as follows. Take $N_\theta$ to be the number of images, $N_\mathrm{s}$ the number of sources, and let $N_\mathrm{c}$ denote the number of mass cells. Then, the linearized problem takes the form

$$\boldsymbol{\phi} = \Gamma \mathbf{x}, \tag{39}$$

where $\Gamma$ is a matrix of dimension $2(N_\theta) \times (N_\mathrm{c} + 2N_\mathrm{s})$, and generally not invertible. $\boldsymbol{\phi}$ is a vector containing position and shear information, whereas $\mathbf{x}$ is a vector containing all the unknowns. A weakness of the work is that it assumes the source galaxies to be point-like.[18] This is, however, corrected for in an a-posteriori manner, by allowing for residuals $\mathbf{r}$ in the lens equation; $\mathbf{r} = \phi - \Gamma\mathbf{x}$. Approximate solutions to the problem are sought by maximizing the functional $\mathcal{L} = \mathrm{e}^{\frac{1}{2}\mathbf{x}^2}$, assuming $\mathbf{r}$ is Gaussian. The $\chi^2$ is now defined as

$$\chi^2 = \mathbf{r}^\mathrm{T} \boldsymbol{C}^{-1} \mathbf{r} \tag{40}$$

where $\boldsymbol{C}$ is the covariance matrix of $\mathbf{r}$. As with most studies combining weak and strong, this work only includes shear.

Since the approach is nonparametric, a regularization procedure is necessary. In Paper I [114], a bi-conjugate gradient method minimizing $\chi^2$, stopping at a target value $\epsilon$, is found useful (implicitly regularizing the problem). Paper III shows that combining weak and strong lensing makes the choice of $\epsilon$ less relevant. Also, since the problem formulation consists in minimizing a quadratic function with linear constraints, a quadratic-programing approach (QADP) is taken. By using a minimization algorithm that is constrained to minimize $\chi^2$ for positive masses, unphysical negative masses are avoided, and the problem is more properly regularized. An application using strong lensing is found in their Paper II [115]. The result of their analysis is a software package known as WSLAP,[19] which is made available for public use. Also, note Diego [116].

The above works illustrate the capability of nonparametric methods to compete with parametric ones, when a properly regularized maximum-likelihood approach is taken.

### 4.5 Using critical curves

As discussed in Sect. 3.3, strong-lensing observables include arcs, which form around critical curves ($\det \mathcal{A} = 0$). This is used in Cacciato et al. [117] to develop a cluster potential reconstruction method relying on critical-curve constraints imposed near arc-forming regions. Building on the least-$\chi^2$ minimization procedure developed in Seitz et al. [30], Bartelmann et al. [24], this becomes a different route to combining weak and strong lensing. Also, note that the weak-lensing constraints are imposed on a low-resolution grid, whereas the strong-lensing constraints are imposed on a more fine-grained grid, so as to concentrate the analysis to the relevant region.

One advantage of this method is the elimination of the need to finding multiple images of the same source, which requires work. Additionally, one ought also know whether or not there are additional sources not imaged. But inverting the lens equation to find out is of course a highly nonlinear process. Furthermore, multiple imaging is caused by sources inside or close to caustics, which limits the applicability of the method.

In Cacciato et al (2006a), measured ellipticities are compared to $\gamma$ rather than to the reduced shear $\gamma(1-\kappa)^{-1}$. This approximation is fixed in the observational follow-up [118], which also improves the grinding of the data and adds a regularization term to not overfit noise. Note also the follow-up [119], which incorporates a mesh-free, free-form approach to lens-mass reconstruction, through the use of radial basis functions. When discretizing the data, a regular mesh is often employed. But, as the authors argue, astronomical data are often quite irregular. The work extends their own previous work, and other, like [24, 30, 120] into the flexible, numerical mesh-free domain.

[18] The paper includes a remark about a follow-up work on this issue, which we have not been able to find.

[19] Weak & Strong Lensing Analysis Package.

### 4.6 Particle-based lensing (PBL)

Let us also make mention of the works by Deb et al. [121, 122], where a particle-based approach is taken. Whereas grid-based techniques are often optimized to measure at a single scale, reality is that the structure of a cluster lens is hierarchical. Thus, a method capable of increased resolution in regions high on information would be desirable. Second, the smoothing and weighting are arguably *ad hoc* by construction (motivated by model and theory), whereas an ideal smoothing scale should be variable. Another point is the abrupt switch in image parity at the critical curves, leading to a discontinuity in the ellipticity as a function of $\kappa$ and $\gamma$. This prevents linear minimization schemes from converging to a 'strong lens'—solution if starting off by a 'weak lens' initial guess.

PBL methods have the ability to combine disparate lensing scales in a coherent way without regularization schemes and are already in use in other fields of astrophysics where a wide range of physical scales are present, such as numerical $N$-body simulations. PBL is a new way of discretizing and describing a reconstructed field, including a metric comparison to the observed data, but PBL is nevertheless not a minimization scheme. PBL consists of a list of potentials (one at each observed lensed image) and a metric to describe the goodness of fit. A $\chi^2$-minimization to estimate a maximum-likelihood potential field is used, and—for practical implementations anyway—some sort of minimization procedure has to be chosen.

Note that the (parameter-free) PBL method developed may in principle combine both weak and strong constraints, including flexion. Applying the technique to the much studied cluster CL-A1689 reveals a secondary mass peak in the north-east direction, confirming previous optical observations (Deb et al. [122]). The study is reported to be the first nonparametric strong and weak (shear) mass map with a covariance matrix. Moreover, refer to Deb et al. [123] for another application of PBL to lensing, and for more on the covariance matrix.

### 4.7 Including flexion in the combined analysis

Including flexion, the lensing equation may be written as [59]

$$\hat{\beta} = \theta - g\theta^* - \frac{1}{4}\Psi_1^*\theta^2 - \frac{1}{2}\Psi_1\theta\theta^* - \frac{1}{4}\Psi_3(\theta^*)^2, \tag{41}$$

where $\theta$ is the complexified image-plane angular coordinate and $^*$ denotes complex conjugation. Note that $\hat{\beta} = \beta/(1-\kappa)$ is the reduced source position, $g$ is the reduced shear (17), and that $\Psi_1 = \frac{\mathcal{F}}{1-\kappa}$ and $\Psi_3 = \frac{\mathcal{G}}{1-\kappa}$ are reduced flexion variables (cf. Eq. (32)), all fit to taking the mass-sheet degeneracy into account.

It is in fact notable that weak lensing including flexion terms has also been combined with strong-lensing constraints in observational studies, as described in this section. First, note the study by Leonard et al. [58], which incorporates strong-lensing constraints with the shapelets formalism developed in the aforementioned study by Goldberg and Bacon [52] for second-order weak lensing. The machinery is applied to the cluster Abell1689. By use of flexion data alone, and with a parametric approach, the study reveals a substructure offset from the central cluster members that was not visible from shear measurements alone. The same is suggested when a nonparametric approach is taken, combining weak- and strong-lensing data. Several studies show the same; flexion is particularly sensitive to cluster substructures (e.g. Bacon et al. [124]). This is perhaps not so surprising from a theoretical point of view, considering the fact that flexion encodes derivatives of the shear map.

The mass reconstruction method presented by Bradač and discussed earlier is extended by Er et al. [63], where a flexion term $\chi_f^2$ is added, reading

$$\sum_{i=1}^{N_\mathrm{f}} \frac{|t_{1i} - G_1(\boldsymbol{\theta}_i, z_i)|^2}{\sigma_{t1}^2} + \frac{|t_{3i} - G_3(\boldsymbol{\theta}_i, z_i)|^2}{\sigma_{t3}^2}, \tag{42}$$

where $t_{1i}$ (or $t_{3i}$) refers to first (or third)-flexion data, and $G_1$ (or $G_3$) to the theoretical model flexion. $N_f$ is the number of images with flexion data. The approach is tested on simulations and concludes that flexion is sensitive to substructures.

Also, note the analytic image model (AIM) method, developed by Cain et al. [59], which is distinct from shapelets- and moment-based methods. Starting from the conservation of surface brightness $I_{\mathrm{obs}}[\theta] = I_{\mathrm{src}}[\beta(\theta)]$, it minimizes the figure of merit given by

$$\chi^2 = \sum_n \frac{(I_{\mathrm{obs}}(\theta^{(n)}) - I_{\mathrm{AIM}}(\theta^{(n)}; p_{\mathrm{int}}, p_{\mathrm{lens}})^2)}{\sigma_n^2}, \tag{43}$$

where $\theta^{(n)}$ is the image-plane position of the $n$th pixel and $\sigma_n^2$ is an estimate of the variance in that pixel's value. Note that $p_{\mathrm{int}}$ and $p_{\mathrm{lens}}$ are arrays containing the parameters of the theory. In particular, $p = \{g_1, g_2, \Psi_{11}, \Psi_{12}, \Psi_{31}, \Psi_{32}\}$, where $g_1$, $\Psi_{11}$ and $\Psi_{31}$ are the real parts ($g_2$, $\Psi_{12}$ and $\Psi_{32}$ are the imaginary parts) of $g$, $\Psi_1$ and $\Psi_3$, respectively. Instead of estimating derived quantities, such as shapelet coefficients or surface brightness moments, the AIM method instead fits a mass-sheet transformation invariant model to each galaxy image. The method is applied to CL–A1689, detecting mass structures. In a follow-up work by Cain et al. [60], the AIM method is applied to clusters, revealing that galaxy-group scale substructures may remain undetected, such as to underestimate the mass in such regions by $\sim 25 - 40$ per cent. With flexion included, subhalo masses of $\sim 3 \times 10^{12} M_\odot$ are detectable at an angular resolution smaller than 10 arcsec.

Albeit not including strong-lensing events per say, we also mention Lanusse et al. [78], which combines weak-lensing shear and flexion data through developing a method that avoids any binning of the irregularly sampled fields. As a result, the mass-mapping problem becomes an ill-posed inverse problem. This is regularized using a robust multi-scale wavelet sparsity prior. Fast Fourier estimators make the algorithm efficient, and the method is tested on simulated data. The conclusion is again that flexion increases the ability to detect substructures.

### 4.8 Combining lensing information with other sources

In this paper, we have focused on gravitational lensing as a probe on galaxy clusters. But, there are other probes as well, which may be summarized as follows:

- Weak- and strong-lensing constrains the line-of-sight projection of the cluster potential
- X-ray observations constrain the density and the temperature of the intracluster gas
- Sunyaev–Zel′dovich effect constrains the gas pressure
- Gradient of gravitational potential can be constrained by galaxy kinematics
- Cluster kinematics. Consider, for instance, the work by Pizzuti et al. [125], where kinematics measurements of galaxy cluster MACS J1206 is combined with weak- and strong-lensing constraints. Also, consider the work by Umetsu et al. [126], which is chosen for further discussion in Section 4.9.

A complete treatment would of course have to combine from all the above approaches. Although alternative sources for information are beyond the scope of the current review, we take the opportunity to mention that work has been done in combining the available information. One example is that of Konrad et al. [127], where reconstruction from thermal gas (X-ray) and gravitational lensing is considered in a nonparametric manner. The ultimate goal of this study is reported to be a nonparametric method combined strong and weak lensing, observations of thermal gas physics and galaxy kinematics into one consistent model for the projected cluster potential. Other examples include that of [128], where lensing information is combined with the thermal Sunyaev–Zel′dovich effect and Tchernin et al. [129, 130], where X-ray and the Sunyaev–Zel′dovich effect are used to reconstruct the gravitational potential by joint analysis. Combining lensing data with the thermal Sunyaev–Zel′dovich effect allows for constraining the mass of clusters without having to make strong assumptions about their hydrodynamical states (cf. Bocquet et al. [131]).

### 4.9 Software developments

As far as software goes, the recent work by Niemiec et al. [132] should be noted, where the combination of weak and strong constraints is implemented in the publically available modelling software package LENSTOOL, under the name *hybrid*-LENSTOOL. The name derives from the process: this package uses a parametric approach to model the core of the cluster and a nonparametric (free-form) approach for the outskirts. The optimization is based on a joint optimization of both weak and strong constraints, such that the combined likelihood function is

$$\mathcal{L}(\boldsymbol{\Theta}, \boldsymbol{w}) = \mathcal{L}_{\mathrm{SL}}(\boldsymbol{\Theta}, \boldsymbol{w}) \times \mathcal{L}_{WL}(\boldsymbol{\Theta}, \boldsymbol{w}). \tag{44}$$

Here, $\boldsymbol{\Theta}$ are the free parameters of the parametric part, whereas the grid model composed of $N$ radial basis functions is ordered in a vector $\boldsymbol{w}$. The details of the strong- and weak-lensing likelihoods are further described in the paper. As a result, the package optimizes weak and strong lensing simultaneously, as opposed to successively, as does the forerunner *Sequential-Fit*. It should be noted, however, that Niemiec et al. [133] still end up using sequential fit, since the strong-lensing constraints optimize the likelihood in the source plane, as opposed to the image plane. This might not be accurate enough for complex clusters, given that best-fit mass reconstruction is the goal. Testing both methods on simulated clusters is left for future work. The recent work by Patel et al. [134] also seems to use LENSTOOL (the hybrid package), although the paper is not specific about it. In this latter paper, it is the cluster MACS J1423 that is under scrutiny. As before, a parametric model is used for the core, where SL constraints are available, whereas a free-form approach is adopted for the WL region (the outskirts).

Another software is SWUnited, based on the Bradač–Hoag model. This algorithm based on the works by Bartelmann et al. [30] (Sect. 3.2) and Bradac et al. [135] (Sect. 4.2) is nonparametric (free-form) and reconstructs the lensing potential (as opposed to the convergence). This means that assumptions on the mass density outside the field of observation are not necessary. Also, the formalism includes flexion constraints.

Note also WSLAP+, which builds on work by Diego et al. and Sendra et al. (discussed in Sect. 4.4) and Lam. This method unites strong and weak (shear) constraints. In Diego et al. [136], critical curves are added to the strong constraints as well. This code is actually used in the Euclid *Early release observations* [137, 138], where the combined analysis of weak- and strong-lensing data gives a unique opportunity to constrain the viral mass, also recovering a solution consistent with the expected NFW profiles of Abell 2390, with $M_{200} = (1.48 \pm 0.29) \times 10^{15} M_{\odot}$.

A genetic algorithm (GA) for nonparametric strong-lensing systems (GRALE) was also developed (Liesenborgs et al. [139], Liesenborgs et al. [140]) and later extended to take time-domain and weak-lensing information into account (Liesenborgs et al. [141]). This method resembles a regularization procedure but is different in that it does not optimize the different criteria together. Instead, a GA looks for a common optimum for the different so-called fitness measures. The results as regards the addition of weak lensing to GRALE [20] are reported to be somewhat mixed, and the procedure is plagued by the mass-sheet degeneracy.

Finally, note also the newly developed RELENSING package [142], written in Python. It is a free-form approach on an adaptive, irregular grid, combining strong and weak (shear) constraints and is tested on the simulated clusters ARES and HERA.

A full analysis and comparison of the different packages available shall not be undertaken here, but we refer the interested reader to Meneghetti et al. [143], where different groups have participated in the reconstruction of simulated clusters, for comparison purposes. Other tests include the Shear TEsting Programme (STEP) and GRavitational LEnsing Accuracy Testing (GREAT).

Finally, also note CLUMI+, a self-consistent, multi-probe methodology for reconstructing the mass distribution in and around galaxy clusters by combining gravitational lensing and dynamical observation (Umetsu et al. [126]). This algorithm builds on the joint-likelihood framework presented in Umetsu [111] and incorporates weak-lensing shear and magnification data with projected escape velocity measurements in the cluster in-fall region, yielding tighter constraints on the gravitational potential without relying on equilibrium assumptions.

#### *4.9.1 The use of artificial intelligence*

As a final remark, it is worthwhile pausing to ponder the future. As numerous authors point out, the amount and quality of the lensing data has been steadily increasing. As a result, one ought only expect that the need for fast, automated analysis becomes more and more prevalent. To this end, people have looked to machine learning, to see if this tool is fit for the job of automating the process of lens-mass reconstruction. Novel attempts have been made, such as that by Hezaveh et al. [144], Galan et al. [145] and [146], all concerning strong lenses. To the best of knowledge, no other attempts at combining weak and strong lensing in an automated ML approach have been reported, besides our own humble attempt at using the roulette formalism to such ends [75, 76], and the recent work on the so-called MARS[21] algorithm due to Cha et al. [147], where a deep-learning technique is applied to navigate the extremely large space of free parameters. The data set of this latter study is obtained from wide-field JWST, which makes it all the more interesting. The algorithm was also applied to a recent free-form reconstruction of the Bullet cluster (1E 0657-56) [148]. By using the algorithm, the assumption of mass following light is alleviated. This is crucial in a complex merger like the Bullet cluster, which is a hard nut to crack in terms of simulation.

There is also another code, MrMARTIAN (Multi-resolution MAximum-entropy Reconstruction Technique Integrating Analytic Node), adapted from MARS, which is useful if one wants to combine a free-form and parametric approach. This comes in handy whenever the SL constraints are to scarce to constrain the potential properly. This algorithm was used in the study of the most distant candidate star ($z \sim 6$) [149–151], WHL0137-LS, nicknamed Earendel, which is hosted within the strongly lensed Sunrise Arc produced by the gravitational potential of the galaxy cluster WHL J013719.8–08284. Its apparent magnitude would require extreme lensing magnification (a few thousand) and exceptionally high luminosity. However, in a recent study by Scofield et al. [152], a joint WL and SL analysis shows that the magnification is probably not more than 40-70. If that is correct, Earendel is not a star. This conclusion might be problematic on its own, since the proposed model does not explain the symmetry observed. Moreover, mass estimates of the lensing cluster are in tension with those of the Planck observation, which was based on the Sunyaev–Zel′dovich effect. This is likely to be due to systematic bias effects, such as an inherent triaxial structure of the cluster. This shows the need for an accurate modelling of clusters.

Finally, before ending, we also mention here a recent study by [153], which uses a gravitationally constrained neural field to reconstruct 3D maps of dark-matter distributions in the Universe from shear fields. After all, such 3D maps are a much sought goal in mass reconstruction, and incorporating SL constraints might be a future step.

## 5 Conclusion and future outlook

In this review, we have traced the development of joint analysis of weak and strong gravitational lensing data, with a particular emphasis on cluster lens-mass reconstruction. In summary, there seems to be two important strands of development.

*First*, there is the development of the theory of weak lensing to account for higher-order terms, such as flexion, second-flexion and terms beyond (so-called roulette expansion). This is a natural extension of the early development of nonlinear methods (such as KS). By including terms beyond shear in the lens map, a bridge is made from the weak regime towards the strong. As of today, the roulette expansion developed by Clarkson is the most complete work to this end, but even here only leading-order screen-space terms have been kept, whereas the rest, such as line-of-sight effects, are wittingly swept under the carpet.

*Second*, with the development of so-called inverse methods (and most prominent among them the class of maximum-likelihood approaches) around the turn of the millennium, the two regimes have arguably been merged as far as statistical data analysis is concerned. The merger of the regimes has proven useful, for instance, in analysing the *Early-Release Observations* of Euclid.

[20] Acronym for GRAvitational LEnses.

[21] Acronym for MAximum-entropy ReconStruction.

The success of this approach hinges on the linearity of the method; that is, it allows for inclusion of strong-lensing effects as linear constraints. This method should therefore also in principle be applicable to the grander project of incorporating also other sources of information, such as time-delay, cluster kinematics constraints, X-ray observations and Sunyaev–Zel′dovich effect constraints in the lens-mass reconstruction process. Such projects should be undertaken in order to accommodate a more accurate understanding of the complex internal three-dimensional structure of clusters.

### 5.1 Future work

A number of improvements are possible. For instance, degeneracies such as the mass-sheet degeneracy still plague the field. One way to resolve this is through several redshifts. Research into other methods, such as adaptive grids and alternative sources of information to break the degeneracy, should be furthered.

In inverse methods using weak-lensing data, inclusion of flexion terms has already proven useful, but terms beyond this order have not been investigated, except theoretically in the aforementioned roulette formalism due to Clarkson, or equivalently, through a strong-lensing approach by Fleury et al. Foreseeing an increase in data quality, such decomposition of image information beyond flexion could prove valuable.

Flexion studies, adaptive grids and combined analysis of weak and strong lensing have revealed substructures in cluster-mass distributions. The development of three-dimensional parametric models of clusters should therefore go on, reaching beyond, for example, the NFW. To such ends, studying the line-of-sight effects from the internal structure of the lens seems to be a natural step. Furthermore, the development of nonparametric methodologies for data fitting must continue, adapting to the increase in both data quality and quantity.

To this end, the recent developments within automated processes by way of machine learning seem like a particularly promising route. Here, we mention the MARS algorithm, which stands out as the most promising candidate of merging the two regimes at the moment. Armed with a formalism capable of drawing information beyond shear, this might be just what we need. As a final point, the seeming scarcity of the literature on the use of machine learning for lens-mass reconstruction is surprising. A review forcing the literature to speak on this point would be in place. Or is the silence a result of failed, and thus unpublished, attempts?

**Acknowledgements** We thank Ms Elise Lindegaard Hanssen, Mr Sangjun Cha, Professor Jose Maria Diego, Dr. Jenny Wagner and Professor Keiichi Umetsu for valuable comments on the manuscript.

**Funding** Open access funding provided by NTNU Norwegian University of Science and Technology (incl St. Olavs Hospital - Trondheim University Hospital).

## Appendix: A study selection

Refer to Table 3 for a listing of the 32 shortlisted papers in this study.

**Table 3** The table presents the shortlist of 32 papers selected for study through the structured searchThe column 'MSD' categorizes how the mass-sheet degeneracy is treated in the study. '$z$' means sources at different redshifts. 'magn' is short for magnification. ' $\leq$ Flex' means that both shear and flexion data have been used

| Authors | Title | Free-form? | $\psi$ or $\kappa$? | MSD | WL data |
|---|---|---|---|---|---|
| AbdelSalam et al. [97] | Mass reconstruction of combined strong & weak lensing | Yes | $\kappa$ | z | Shear |
| AbdelSalam et al. [99] | Nonparametric reconstruction of Abell 2218 from combined weak and strong lensing | Yes | $\kappa$ | z | Shear |
| AbdelSalam et al. [96] | Cluster mass profile from lensing | Yes | $\kappa$ | z | Shear |
| Birrer et al. [70] | Line-of-sight effects in strong lensing: putting theory into practice | No | $\kappa$? | Discussed | Shear |
| Bradac [104] | Strong and weak lensing united: the cluster mass distribution of the most X-ray | Yes | $\psi$ | z | Shear |
| | luminous cluster RX J1347-1145 | | | | |
| Bradac et al. [107] | Strong & weak lensing united: The cluster mass distribution of RX J1347-1145 | Yes | $\psi$ | z | Shear |
| Bradač et al. [103] | Strong and weak lensing united—I. The combined strong and weak lensing | Yes | $\psi$ | z | Shear |
| | cluster mass reconstruction method | | | | |
| Bradač et al. [103] | Strong and weak lensing united—II. The cluster mass distribution of the most | Yes | $\psi$ | z | Shear |
| | X-ray luminous cluster RX J1347.5-1145 | | | | |
| Bradač et al. [103] | Strong and weak lensing united—III. Measuring the mass distribution of the merging | Yes | $\psi$ | z | Shear |
| | galaxy cluster 1ES 0657-558 | | | | |
| Bradač et al. [154] | Focusing cosmic telescopes: exploring redshift z 5-6 galaxies with the bullet cluster | Yes | $\psi$ | z | Shear |
| | 1E0657-56 | | | | |
| Cacciato et al. [117] | Combining weak and strong lensing in cluster potential reconstruction | Yes? | $\psi$ | Scale | Shear |
| Cha et al. [147] | Precision MARS Mass Reconstruction of A2744: Synergizing the Largest | Yes | $\kappa$ | Photometric | Shear |
| Strong-lensing and Densest Weak-lensing Data Sets from JWST | | | | | |
| Cha et al. [148] | A High-Caliber View of the Bullet Cluster Through JWST Strong and Weak Lensing Analyses | Yes | $\kappa$ | Photometric | Shear |
| Deb et al. [122] | Mass reconstruction using Particle Based Lensing II: Quantifying substructure with | Yes | $\psi$ | z(+ scale?) | Shear |
| | strong + weak lensing and X-rays | | | | |
| Diego [116] | An improved dark matter map from weak and strong lensing in A1689 | Yes | $\kappa$ | Discussed | Shear |
| Diego et al. [113] | Combined weak and strong lensing data with WSLAP | Yes | $\kappa$ | Discussed | Shear |
| Er et al. [63] | Mass reconstruction by gravitational shear and flexion | Yes | $\psi$ | z? | $\leq$ Flex. |

**Table 3** continued

| Authors | Title | Free-form? | $\psi$ or $\kappa$? | MSD | WL data |
|---|---|---|---|---|---|
| Leonard et al. [58] | Gravitational shear, flexion and strong lensing in Abell 1689 | Partially | $\kappa$ | Discussed | $\leq$ Flex. |
| Liesenborgs et al. [141] | Extended lens reconstructions with grale: exploiting time-domain, substructural and weak lensing information | Yes | $\kappa$ | Discussed | Shear |
| Merten et al. [118] | Combining weak and strong cluster lensing: applications to simulations and MS 2137 | Yes | $\psi$ | Discussed | Shear |
| Authors | Title | Free-form? | $\psi$ or $\kappa$? | MSD | WL data |
| Merten [119] | Mesh-free free-form lensing - I. Methodology and application to mass reconstruction | Yes | $\psi$ | Discussed | Shear |
| Natarajan et al. [112] | The survival of dark matter halos in the cluster CI0024+ 16 | No | $\kappa$ | Not discussed | Shear |
| Niemiec et al. [132] | Hybrid-lenstool: a self-consistent algorithm to model galaxy clusters with strong- and weak-lensing simultaneously | Partially | $\kappa$ | Not discussed | Shear |
| Normann and Clarkson [74] | Recursion relations for gravitational lensing | Yes | $\psi$ | Not discussed | $n$th order |
| Niemiec et al. [133] | Beyond the ultradeep frontier fields and legacy observations (BUFFALO): a high-resolution strong+weak lensing view of Abell 370 | Partially | $\kappa$ | Not discussed | Shear |
| Patel et al. [134] | The KALEIDOSCOPE survey: a new strong and weak gravitational lensing view of the massive galaxy cluster MACS J1423.8+2404 | Partially | $\psi$ | Not discussed | Shear |
| Ponente and Diego [90] | Systemics in lensing reconstruction: dark matter rings in the sky? | Yes | $\kappa$ | $z$ | Shear |
| Saha et al. [98] | Cluster reconstruction from combined strong and weak lensing | Yes | $\kappa$ | z | Shear |
| Schneider and Seitz [16] | Steps towards nonlinear cluster inversion through gravitational distortions. I. Basic considerations and circular clusters | Yes | $\kappa$ | Discussed | Shear |
| Scofield et al. [152] | Is Earendel a Star?: Investigating the Sunrise Arc Using JWST Strong and Weak Gravitational Lensing Analyses | Partially | $\kappa$ | Not discussed | Shear |
| Umetsu [111] | Model-free multi-probe lensing reconstruction of cluster mass profiles | Yes | $\kappa$ | $z$ and magn | Shear |
| Umetsu and Broadhurst [108] | Combining lens distortion and depletion to map the mass distribution of A1689 | Yes | $\kappa$ | $z$ and magn | Shear |

## References

1. I. Newton, Opticks: Or, A Treatise of the Reflections, Refractions, Inflexions and Colours of Light. Also Two Treatises of the Species and Magnitude of Curvilinear Figures. London : Printed for Sam Smith, and Benj. Walford, printers to the Royal Society, at the prince's Arms in St. paul's Church-yard, ??? (1704). https://archive.org/details/opticksortreatis00newt/page/n6
2. J. Michell, On the means of discovering the distance, magnitude, &c. of the fixed stars, in consequence of the diminution of the velocity of their light, in case such a diminution should be found to take place in any of them, and such other data should be procured from observations, as would be farther necessary for that purpose. by the rev. john michell, b. d. f. r. s. in a letter to henry cavendish, esq. f. r. s. and a. s. Philosophical Transactions of the Royal Society of London (1776-1886) **74** (1784) https://doi.org/10.1098/rstl.1784.0008
3. S.P. Laplace, Exposition du système du monde (1795)
4. P. Schneider, J. Ehlers, E. Falco, Gravitational Lenses vol. 33, p. 560. Springer, Cham (1999). https://doi.org/10.1007/978-3-662-03758-4
5. P. Schneider, C. Kochanek, J. Wambsganss, Gravitational Lensing: Strong, Weak and Micro: Saas-Fee Advanced Course 33 vol. 33, pp. 150–152. Springer, Cham (2006)
6. A. Congdon, C.R. Keeton, Principles of Gravitational Lensing, 1st edn., p. 287. Springer, Cham (2018). https://doi.org/10.1007/978-3-030-02122-1
7. K. Umetsu, Cluster-galaxy weak lensing. Astron. Astrophys. Rev. **28**(1), 7 (2020). https://doi.org/10.1007/s00159-020-00129-w. arXiv:2007.00506 [astro-ph.CO]
8. A. Morandi, K. Pedersen, M. Limousin, Reconstructing the triaxiality of the galaxy cluster a1689: solving the x-ray and strong lensing mass discrepancy. Astrophys. J. **729**(1), 37 (2011)
9. D.J. Bacon, B.M. Schafer, Twist and turn: weak lensing image distortions to second order. Mon. Not. R. Astron. Soc. **396**(4), 2167–2175 (2009). https://doi.org/10.1111/j.1365-2966.2009.14850.x
10. J.A. Tyson, F. Valdes, R.A. Wenk, Detection of systematic gravitational lens galaxy image alignments - mapping dark matter in galaxy clusters. Astrophys. J. **349**(1), 1 (1990). https://doi.org/10.1086/185636
11. M. Jauzac, E. Jullo, J.-P. Kneib, H. Ebeling, A. Leauthaud, C.-J. Ma, M. Limousin, R. Massey, J. Richard, A weak lensing mass reconstruction of the large-scale filament feeding the massive galaxy cluster macs j0717. 5+ 3745. Mon. Not. R. Astron. Soc. **426**(4), 3369–3384 (2012)
12. N. Kaiser, G. Squires, Mapping the dark matter with weak gravitational lensing. Astrophys. J. **404**, 441–450 (1993). https://doi.org/10.1086/172297
13. J. Miralda-Escude, The correlation-function of galaxy ellipticities produced by gravitational lensing. Astrophys. J. **380**(1), 1–8 (1991). https://doi.org/10.1086/170555
14. J. Miralda-Escude, Gravitational Lensing by Clusters of Galaxies: Constraining the Mass Distribution. Astrophys. J. **370**, 1 (1991). https://doi.org/10.1086/169789
15. M.A. Troxel, M. Ishak, The intrinsic alignment of galaxies and its impact on weak gravitational lensing in an era of precision cosmology. Phys. Rep. **558**, 1–59 (2015). https://doi.org/10.1016/j.physrep.2014.11.001. arXiv:1407.6990 [astro-ph.CO]
16. P. Schneider, C. Seitz, Steps towards nonlinear cluster inversion through gravitational distortions. I. Basic considerations and circular clusters. Astron. Astrophys. **294**, 411–431 (1995). https://doi.org/10.48550/arXiv.astro-ph/9407032. arXiv:astro-ph/9407032 [astro-ph]
17. C. Seitz, P. Schneider, Steps towards nonlinear cluster inversion through gravitational distortions II Generalization of the Kaiser and Squires method. Astron. Astrophys. **297**, 287 (1995). https://doi.org/10.48550/arXiv.astro-ph/9408050. arXiv:astro-ph/9408050 [astro-ph]
18. P. Schneider, Cluster lens reconstruction using only observed local data. Astron. Astrophys. **302**(3), 639–648 (1995)
19. S. Seitz, P. Schneider, Cluster lens reconstruction using only observed local data: an improved finite-field inversion technique. Astron. Astrophys. **305**, 383 (1996). https://doi.org/10.48550/arXiv.astro-ph/9503096. arXiv:astro-ph/9503096 [astro-ph]
20. N. Kaiser, G. Squires, T. Broadhurst, A method for weak lensing observations. Astrophys. J. **449**(2), 460–475 (1995). https://doi.org/10.1086/176071
21. M. Bartelmann, Cluster mass estimates from weak lensing. Astron. Astrophys. **303**, 643 (1995). https://doi.org/10.48550/arXiv.astro-ph/9412051. arXiv:astro-ph/9412051 [astro-ph]
22. C. Seitz, P. Schneider, Steps towards nonlinear cluster inversion through gravitational distortions. III. Including a redshift distribution of the sources. Astron. Astrophys. **318**, 687–699 (1997). https://doi.org/10.48550/arXiv.astro-ph/9601079. arXiv:astro-ph/9601079 [astro-ph]
23. S. Seitz, P. Schneider, A new finite-field mass reconstruction algorithm. Astron. Astrophys. **374**, 740–745 (2001). https://doi.org/10.1051/0004-6361:20010493
24. S. Seitz, P. Schneider, M. Bartelmann, Entropy-regularized maximum-likelihood cluster mass reconstruction. Astron. Astrophys. **337**, 325–337 (1998). https://doi.org/10.48550/arXiv.astro-ph/9803038. arXiv:astro-ph/9803038 [astro-ph]
25. M. Lombardi, G. Bertin, Improving the accuracy of mass reconstructions from weak lensing: local shear measurements. Astron. Astrophys. **330**, 791–800 (1998). https://doi.org/10.48550/arXiv.astro-ph/9704219. arXiv:astro-ph/9704219 [astro-ph]
26. M. Lombardi, G. Bertin, Improving the accuracy of mass reconstructions from weak lensing: from the shear map to the mass distribution. Astron. Astrophys. **335**, 1–11 (1998). https://doi.org/10.48550/arXiv.astro-ph/9801244. arXiv:astro-ph/9801244 [astro-ph]
27. M. Lombardi, P. Schneider, C. Morales-Merino, The noise of cluster mass reconstructions from a source redshift distribution. Astron. Astrophys. **382**(3), 769–786 (2002). https://doi.org/10.1051/0004-6361:20011691
28. M. Lombardi, G. Bertin, A fast direct method of mass reconstruction for gravitational lenses. Astron. Astrophys. **348**, 38–42 (1999). https://doi.org/10.48550/arXiv.astro-ph/9906115. arXiv:astro-ph/9906115 [astro-ph]
29. G. Squires, N. Kaiser, Unbiased cluster lens reconstruction. Astrophys. J. **473**(1), 65–80 (1996). https://doi.org/10.1086/178127
30. M. Bartelmann, R. Narayan, S. Seitz, P. Schneider, Maximum-likelihood cluster reconstruction. Astrophys. J. Lett. **464**, 115 (1996). https://doi.org/10.1086/310114. arXiv:astro-ph/9601011 [astro-ph]
31. S.L. Bridle, M.P. Hobson, A.N. Lasenby, R. Saunders, A maximum-entropy method for reconstructing the projected mass distribution of gravitational lenses. Mon. Not. R. Astron. Soc. **299**(3), 895–903 (1998)
32. P. Marshall, Physical component analysis of galaxy cluster weak gravitational lensing data. Mon. Not. R. Astron. Soc. **372**(3), 1289–1298 (2006)
33. M. Bartelmann, P. Schneider, Weak gravitational lensing. Phys. Rep. **340**(4–5), 291–472 (2001). https://doi.org/10.1016/S0370-1573(00)00082-X. arXiv:astro-ph/9912508 [astro-ph]
34. E. Falco, M. Gorenstein, I. Shapiro, On model-dependent bounds on h (0) from gravitational images application of q0957+ 561a, b. Astrophys. J. **289**, 1–4 (1985)
35. M.V. Gorenstein, E.E. Falco, I.I. Shapiro, Degeneracies in parameter estimates for models of gravitational lens systems. Astrophys. J. **327**, 693 (1988). https://doi.org/10.1086/166226
36. P. Saha, Lensing degeneracies revisited. Astron. J. **120**(4), 1654–1659 (2000). https://doi.org/10.1086/301581. arXiv:astro-ph/0006432 [astro-ph]
37. J. Liesenborgs, S. De Rijcke, H. Dejonghe, P. Bekaert, A generalization of the mass-sheet degeneracy producing ring-like artefacts in the lens mass distribution. Mon. Not. R. Astron. Soc. **386**(1), 307–312 (2008)

38. P. Saha, L.L.R. Williams, Gravitational lensing model degeneracies: is steepness all-important? Astrophys. J. **653**(2), 936–941 (2006). https://doi.org/10.1086/508798. arXiv:astro-ph/0608496 [astro-ph]
39. J. Liesenborgs, S. De Rijcke, Lensing degeneracies and mass substructure. Monthly Notices RAS **425**(3), 1772–1780 (2012). https://doi.org/10.1111/j.1365-2966.2012.21751.x. arXiv:1207.4692 [astro-ph.CO]
40. P. Schneider, D. Sluse, Source-position transformation: an approximate invariance in strong gravitational lensing. Astron. Astrophys. **564**, 103 (2014). https://doi.org/10.1051/0004-6361/201322106
41. J. Wagner, Generalised model-independent characterisation of strong gravitational lenses - vi. the origin of the formalism intrinsic degeneracies and their influence on h0. Mon. Not. R. Astron. Soc. **487**(4), 4492–4503 (2019). https://doi.org/10.1093/mnras/stz1587
42. T. Broadhurst, M. Takada, K. Umetsu, X. Kong, N. Arimoto, M. Chiba, T. Futamase, The surprisingly steep mass profile of A1689, from a lensing analysis of Subaru images. Astrophys. J. Lett. **619**(2), 143–146 (2005). https://doi.org/10.1086/428122. arXiv:astro-ph/0412192 [astro-ph]
43. T.J. Broadhurst, A. Taylor, J. Peacock, Mapping cluster mass distributions via gravitational lensing of background galaxies. arXiv preprint arXiv:astro-ph/9406052 (1994)
44. P. Schneider, L. King, T. Erben, Cluster mass profiles from weak lensing: constraints from shear and magnification information. Astron. Astrophys. **353**, 41–56 (2000)
45. M. Bradač, M. Lombardi, P. Schneider, Mass-sheet degeneracy: Fundamental limit on the cluster mass reconstruction from statistical (weak) lensing. Astron. Astrophys. **424**(1), 13–22 (2004)
46. F. Zwicky, Nebulae as gravitational lenses. Phys. Rev. **51**(4), 290–290 (1937). https://doi.org/10.1103/PhysRev.51.290
47. F. Zwicky, On the probability of detecting nebulae which act as gravitational lenses. Phys. Rev. **51**(8), 679–679 (1937). https://doi.org/10.1103/PhysRev.51.679
48. D.M. Goldberg, P. Natarajan, The galaxy octopole moment as a probe of weak-lensing shear fields. Astrophys. J. **564**(1), 65 (2002). https://doi.org/10.1086/324202
49. D.M. Goldberg, A. Leonard, Measuring flexion. Astrophys. J. **660**(2), 1003–1015 (2007). https://doi.org/10.1086/513137
50. Y. Okura, K. Umetsu, T. Futamase, A new measure for weak-lensing flexion. Astrophys. J. **660**(2), 995–1002 (2007). https://doi.org/10.1086/513135
51. Y. Okura, K. Umetsu, T. Futamase, A method for weak-lensing flexion analysis by the HOLICs moment approach. Astrophys. J. **680**(1), 1–16 (2008). https://doi.org/10.1086/587676. arXiv:0710.2262 [astro-ph]
52. D.M. Goldberg, D.J. Bacon, Galaxy-galaxy flexion: Weak lensing to second order. Astrophys. J. **619**(2), 741–748 (2005). https://doi.org/10.1086/426782
53. D.J. Bacon, D.M. Goldberg, B.T.P. Rowe, A.N. Taylor, Weak gravitational flexion. Mon. Not. R. Astron. Soc. **365**(2), 414–428 (2006). https://doi.org/10.1111/j.1365-2966.2005.09624.x
54. J. Irwin, M.V. Shmakova, Observation of cluster substructure using higher order weakly lensed moments. In: American Astronomical Society Meeting Abstracts. American Astronomical Society Meeting Abstracts, vol. 203, pp. 120–06 (2003)
55. J. Irwin, M. Shmakova, Higher moments in weak gravitational lensing and dark matter structures. New Astronomy Reviews 49(2), 83–91 (2005) https://doi.org/10.1016/j.newar.2005.01.032 . Sources and Detection of Dark Matter and Dark Energy in the Universe
56. J. Irwin, M. Shmakova, Observation of small-scale structure using sextupole lensing. Astrophys. J. **645**(1), 17–43 (2006). https://doi.org/10.1086/504100. arXiv:astro-ph/0504200 [astro-ph]
57. J. Irwin, M. Shmakova, J. Anderson, Lensing signals in the hubble ultra deep field using all second-order shape deformations. Astrophys. J. **671**(2), 1182–1195 (2007). https://doi.org/10.1086/522819. arXiv:astro-ph/0607007 [astro-ph]
58. A. Leonard, D.M. Goldberg, J.L. Haaga, R. Massey, Gravitational shear, flexion, and strong lensing in Abell 1689. Astrophys. J. **666**(1), 51–63 (2007). https://doi.org/10.1086/520109. arXiv:astro-ph/0702242 [astro-ph]
59. B. Cain, P.L. Schechter, M.W. Bautz, Measuring gravitational lensing flexion in A1689 using an analytic image model. Astrophys. J. **736**(1), 43 (2011). https://doi.org/10.1088/0004-637X/736/1/43. arXiv:1103.0551 [astro-ph.CO]
60. B. Cain, M. Bradač, R. Levinson, Reconstruction of small-scale galaxy cluster substructure with lensing flexion. Mon. Not. R. Astron. Soc. **463**(4), 4287–4300 (2016). https://doi.org/10.1093/mnras/stw2270
61. A.J. Hawken, S.L. Bridle, Gravitational ' ' ' 'flexion by elliptical dark matter haloes. Monthly Notices of the RAS **400**(3), 1132–1138 (2009). https://doi.org/10.1111/j.1365-2966.2009.15539.x. arXiv:0903.3938 [astro-ph.CO]
62. P. Schneider, X. Er, Weak lensing goes bananas: what flexion really measures. Astron. Astrophys. **485**(2), 363–376 (2008). https://doi.org/10.1051/0004-6361:20078631
63. X. Er, G. Li, P. Schneider, Mass reconstruction by gravitational shear and flexion. arXiv preprint arXiv:1008.3088 (2010)
64. P.D. Lasky, C.J. Fluke, Shape, shear and flexion: an analytic flexion formalism for realistic mass profiles. Mon. Not. R. Astron. Soc. **396**(4), 2257–2268 (2009). https://doi.org/10.1111/j.1365-2966.2009.14888.x
65. C.J. Fluke, P.D. Lasky, Shape, shear and flexion - ii. quantifying the flexion formalism for extended sources with the ray-bundle method. Mon. Not. R. Astron. Soc. **416**(3), 1616–1628 (2011). https://doi.org/10.1111/j.1365-2966.2011.18403.x
66. C.J. Fluke, R.L. Webster, D.J. Mortlock, The ray-bundle method for calculating weak magnification by gravitational lenses. Monthly Notices RAS **306**(3), 567–574 (1999). https://doi.org/10.1046/j.1365-8711.1999.02534.x. arXiv:astro-ph/9812300 [astro-ph]
67. P. Fleury, J. Larena, J.P. Uzan, Weak gravitational lensing of finite beams. Phys. Rev. Lett. (2017). https://doi.org/10.1103/PhysRevLett.119.191101
68. P. Fleury, J. Larena, J.-P. Uzan, Cosmic convergence and shear with extended sources. Phys. Rev. D (2019). https://doi.org/10.1103/physrevd.99.023525
69. P. Fleury, J. Larena, J.-P. Uzan, Weak lensing distortions beyond shear. Phys. Rev. D (2019). https://doi.org/10.1103/physrevd.99.023526
70. S. Birrer, C. Welschen, A. Amara, A. Refregier, Line-of-sight effects in strong lensing: putting theory into practice. J. Cosmol. Astropart. Phys. **2017**(04), 049 (2017)
71. C. Clarkson, The general theory of secondary weak gravitational lensing. J. Cosmol. Astropart. Phys. **2015**(09), 033–033 (2015). https://doi.org/10.1088/1475-7516/2015/09/033
72. C. Clarkson, Roulettes: a weak lensing formalism for strong lensing: I overview. Classical Quantum Gravity (2016). https://doi.org/10.1088/0264-9381/33/16/16lt01
73. C. Clarkson, Roulettes: a weak lensing formalism for strong lensing: Ii. derivation and analysis. Classical and Quantum Gravity (2016) https://doi.org/10.1088/0264-9381/33/24/245003
74. B.D. Normann, C. Clarkson, Recursion relations for gravitational lensing. General Relativity and Gravitation **52**(3) (2020) https://doi.org/10.1007/s10714-020-02677-z
75. H.G. Schaathun, B.D. Normann, E.L. Austnes, S. Ingebrigtsen, S.W. Remøy, S.N. Runde, On the simulation of gravitational lensing. In: Vicario, E., Bandinelli, R., Fani, V., Mastroianni, M. (eds.) Proceedings of the 30th European Conference on Modelling and Simulation. ECMS - European Council for Modelling and Simulation, ??? (2023). Florence, Italy, 21-23 June 2023

76. H.G. Schaathun, B.D. Normann, K. Solevåg-Hoti, Om å kartleggja mørk materie med maskinlæring. Norsk IKT-konferanse for forskning og utdanning (2023). To be presented at *Norsk Informatikkonferanse* in Stavanger 28-29 November 2023
77. J. Vines, Geodesic deviation at higher orders via covariant bitensors. General Relativity and Gravitation **47**(5) (2015) https://doi.org/10.1007/s10714-015-1901-9
78. F. Lanusse, J.-L. Starck, A. Leonard, S. Pires, High resolution weak lensing mass mapping combining shear and flexion. Astron. Astrophys. **591**, 2 (2016). https://doi.org/10.1051/0004-6361/201628278. arXiv:1603.01599 [astro-ph.CO]
79. M. Viola, P. Melchior, M. Bartelmann, Shear-flexion cross-talk in weak-lensing measurements. Monthly Notices of the RAS **419**(3), 2215–2225 (2012). https://doi.org/10.1111/j.1365-2966.2011.19872.x. arXiv:1107.3920 [astro-ph.CO]
80. B. Rowe, D. Bacon, R. Massey, C. Heymans, B. Häußler, A. Taylor, J. Rhodes, Y. Mellier, Flexion measurement in simulations of Hubble Space Telescope data. Monthly Notices of the RAS **435**(1), 822–844 (2013). https://doi.org/10.1093/mnras/stt1353. arXiv:1211.0966 [astro-ph.CO]
81. R.L. Webster, Gravitational lensing and galaxy shape. Monthly Notices of the RAS **213**, 871–888 (1985). https://doi.org/10.1093/mnras/213.4.871
82. C.S. Kochanek, Inverting cluster gravitational lenses. Mon. Not. R. Astron. Soc. **247**(1), 135–151 (1990)
83. R. Lynds, V. Petrosian, Giant luminous arcs in clusters of galaxies. In: Bulletin of the American Astronomical Society **18**, 1014 (1986)
84. G. Soucail, Y. Mellier, B. Fort, F. Hammer, G. Mathez, Further data on the blue ring-like structure in a 370. Astron. Astrophys. **184**, 7–9 (1987)
85. J.F. Navarro, C.S. Frenk, S.D.M. White, Simulations of X-ray clusters. Monthly Notices of the RAS **275**(3), 720–740 (1995). https://doi.org/10.1093/mnras/275.3.720. arXiv:astro-ph/9408069 [astro-ph]
86. J.F. Navarro, C.S. Frenk, S.D.M. White, The structure of cold dark matter halos. Astrophys. J. **462**, 563 (1996). https://doi.org/10.1086/177173
87. J.F. Navarro, C.S. Frenk, S.D.M. White, A universal density profile from hierarchical clustering. Astrophys. J. **490**(2), 493–508 (1997). https://doi.org/10.1086/304888. arXiv:astro-ph/9611107 [astro-ph]
88. A.T. Lefor, T. Futamase, M. Akhlaghi, A systematic review of strong gravitational lens modeling software. New Astron. Rev. **57**(1–2), 1–13 (2013)
89. M.J. Jee, H.C. Ford, G.D. Illingworth, R.L. White, T.J. Broadhurst, D.A. Coe, G.R. Meurer, A. van der Wel, N. Benítez, J.P. Blakeslee, R.J. Bouwens, L.D. Bradley, R. Demarco, N.L. Homeier, A.R. Martel, S. Mei, Discovery of a ringlike dark matter structure in the core of the galaxy cluster Cl 0024+17. Astrophys. J. **661**(2), 728–749 (2007). https://doi.org/10.1086/517498. arXiv:0705.2171 [astro-ph]
90. P.P. Ponente, J.M. Diego, Systematics in lensing reconstruction: dark matter rings in the sky? Astron. Astrophys. **535**, 119 (2011)
91. I. Sendra, J.M. Diego, T. Broadhurst, R. Lazkoz, Enabling non-parametric strong lensing models to derive reliable cluster mass distributions-wslap+. Mon. Not. R. Astron. Soc. **437**(3), 2642–2651 (2014)
92. E. Jullo, J.-P. Kneib, Multiscale cluster lens mass mapping-i. strong lensing modelling. Mon. Not. R. Astron. Soc. **395**(3), 1319–1332 (2009)
93. B. Beauchesne, B. Clément, J. Richard, J.-P. Kneib, Improving parametric mass modelling of lensing clusters through a perturbative approach. Mon. Not. R. Astron. Soc. **506**(2), 2002–2019 (2021)
94. D. Abriola, Combined strong and weak gravitational lensing mass measurements in clusters of galaxies. Proc. Int. Astron. Union **18**(S381), 79–84 (2022). https://doi.org/10.1017/S1743921323004052
95. H.M. AbdelSalam, P. Saha, L.L.R. Williams, Non-parametric reconstruction of cluster mass distribution from strong lensing: modelling abell 370. Mon. Not. R. Astron. Soc. **294**(4), 734–746 (1998). https://doi.org/10.1111/j.1365-8711.1998.01356.x
96. H.M. AbdelSalam, P. Saha, L.L.R. Williams, Cluster Mass Profile from Lensing. arXiv (1997). https://doi.org/10.48550/ARXIV.ASTRO-PH/9710306. arxiv:astro-ph/9710306
97. H.M. AbdelSalam, P. Saha, L.L.R. Williams, Mass reconstruction from combined strong & weak lensing. New Astron. Rev. **42**(2), 157–161 (1998). https://doi.org/10.1016/S1387-6473(98)00040-2. Gravitational Lensing: Nature's Own Weighing Scales
98. P. Saha, L.L. Williams, H. AbdelSalam, Cluster reconstruction from combined strong and weak lensing. arXiv preprint arXiv:astro-ph/9909249 (1999)
99. H.M. AbdelSalam, P. Saha, L.L. Williams, Nonparametric reconstruction of abell 2218 from combined weak and strong lecluster reconstruction from combined strong and weak lensingnsing. Astron. J. **116**(4), 1541 (1998)
100. J.-P. Kneib, P. Hudelot, R.S. Ellis, T. Treu, G.P. Smith, P. Marshall, O. Czoske, I. Smail, P. Natarajan, A wide-field hubble space telescope study of the cluster cl 0024+ 1654 at z= 0.4. ii. the cluster mass distribution. Astrophys. J. **598**(2), 804 (2003)
101. G.P. Smith, J.-P. Kneib, I. Smail, P. Mazzotta, H. Ebeling, O. Czoske, A hubble space telescope lensing survey of x-ray luminous galaxy clusters-iv. mass, structure and thermodynamics of cluster cores at z= 0.2. Mon. Not. R. Astron. Soc. **359**(2), 417–446 (2005)
102. M. Bradač, P. Schneider, T. Erben, M. Lombardi, Strong & weak lensing united: the cluster mass distribution of rx j1347–1145. Proc. Int. Astron. Union **2004**(IAUS225), 155–160 (2004)
103. M. Bradač, P. Schneider, M. Lombardi, T. Erben, Strong and weak lensing united-i. the combined strong and weak lensing cluster mass reconstruction method. Astron. Astrophys. **437**(1), 39–48 (2005)
104. M. Bradac, Strong and weak lensing united: the cluster mass distribution of the most x-ray luminous cluster rxj1347-1145. PoS, 014 (2005)
105. M. Bradač, T. Erben, P. Schneider, H. Hildebrandt, M. Lombardi, M. Schirmer, J.-M. Miralles, D. Clowe, S. Schindler, Strong and weak lensing united-ii. the cluster mass distribution of the most x-ray luminous cluster rx j1347. 5–1145. Astron. Astrophys. **437**(1), 49–60 (2005)
106. M. Bradač, D. Clowe, A.H. Gonzalez, P. Marshall, W. Forman, C. Jones, M. Markevitch, S. Randall, T. Schrabback, D. Zaritsky, Strong and weak lensing united. iii. measuring the mass distribution of the merging galaxy cluster 1es 0657–558. Astrophys. J. **652**(2), 937 (2006)
107. M. Bradac, P. Schneider, T. Erben, M. Lombardi, Strong & weak lensing united: the cluster mass distribution of rx j1347–1145. In: Gravitational Lensing Impact on Cosmology **225**, 155–160 (2005)
108. K. Umetsu, T. Broadhurst, Combining lens distortion and depletion to map the mass distribution of a1689. Astrophys. J. **684**(1), 177 (2008)
109. E. Medezinski, T. Broadhurst, K. Umetsu, D. Coe, N. Benítez, H. Ford, Y. Rephaeli, N. Arimoto, X. Kong, Using weak-lensing dilution to improve measurements of the luminous and dark matter in A1689. Astrophys. J. **663**(2), 717–733 (2007). https://doi.org/10.1086/518638. arXiv:astro-ph/0608499 [astro-ph]
110. K. Umetsu, T. Broadhurst, A. Zitrin, E. Medezinski, L.-Y. Hsu, Cluster Mass Profiles from a Bayesian Analysis of Weak Lensing Distortion and Magnification Measurements: Applications to Subaru Data. Astrophys. J. **729**, 127 (2011). https://doi.org/10.1088/0004-637X/729/2/127. arXiv:1011.3044 [astro-ph.CO]
111. K. Umetsu, Model-free multi-probe lensing reconstruction of cluster mass profiles. Astrophys. J. **769**(1), 13 (2013)
112. P. Natarajan, J.-P. Kneib, I. Smail, T. Treu, R. Ellis, S. Moran, M. Limousin, O. Czoske, The survival of dark matter halos in the cluster cl 0024+ 16. Astrophys. J. **693**(1), 970 (2009)
113. J. Diego, M. Tegmark, P. Protopapas, H. Sandvik, Combined reconstruction of weak and strong lensing data with wslap. Mon. Not. R. Astron. Soc. **375**(3), 958–970 (2007)
114. J. Diego, P. Protopapas, H. Sandvik, M. Tegmark, Non-parametric inversion of strong lensing systems. Mon. Not. R. Astron. Soc. **360**(2), 477–491 (2005)
115. J. Diego, H. Sandvik, P. Protopapas, M. Tegmark, N. Benítez, T. Broadhurst, Non-parametric mass reconstruction of a1689 from strong lensing data with the strong lensing analysis package. Mon. Not. R. Astron. Soc. **362**(4), 1247–1258 (2005)

116. J.M. Diego, An improved dark matter map from weak and strong lensing in a1689. In: Proceedings of the 49th Rencontres de Moriond on Cosmology (2014)
117. M. Cacciato, M. Bartelmann, M. Meneghetti, L. Moscardini, Combining weak and strong lensing in cluster potential reconstruction. Astron. Astrophys. **458**(2), 349–356 (2006)
118. J. Merten, M. Cacciato, M. Meneghetti, C. Mignone, M. Bartelmann, Combining weak and strong cluster lensing: applications to simulations and ms 2137. Astron. Astrophys. **500**(2), 681–691 (2009)
119. J. Merten, Mesh-free free-form lensing-i. methodology and application to mass reconstruction. Mon. Not. R. Astron. Soc. **461**(3), 2328–2345 (2016)
120. M. Bradac, P. Schneider, M. Lombardi, T. Erben, Strong and weak lensing united - i. the combined strong and weak lensing cluster mass reconstruction method. Astron. Astrophys. **437**(1), 39–43 (2005). https://doi.org/10.1051/0004-6361:20042233
121. S. Deb, D.M. Goldberg, V.J. Ramdass, Reconstruction of cluster masses using particle based lensing. i. application to weak lensing. Astrophys. J. **687**(1), 39–49 (2008). https://doi.org/10.1086/590544
122. S. Deb, A. Morandi, K. Pedersen, S. Riemer-Sorensen, D.M. Goldberg, H. Dahle, Mass reconstruction using particle based lensing ii: Quantifying substructure with strong+ weak lensing and x-rays. arXiv preprint arXiv:1201.3636 (2012)
123. S. Deb, D.M. Goldberg, C. Heymans, A. Morandi, Measuring dark matter ellipticity of A901/902 using particle-based lensing. Astrophys. J. **721**(1), 124–136 (2010). https://doi.org/10.1088/0004-637X/721/1/124. arXiv:0912.4260 [astro-ph.CO]
124. D.J. Bacon, A. Amara, J.I. Read, Measuring dark matter substructure with galaxy-galaxy flexion statistics: galaxy-galaxy flexion statistics. Mon. Not. R. Astron. Soc. **409**(1), 389–395 (2010). https://doi.org/10.1111/j.1365-2966.2010.17316.x
125. L. Pizzuti, I.D. Saltas, K. Umetsu, B. Sartoris, Probing vainsthein-screening gravity with galaxy clusters using internal kinematics and strong and weak lensing. Mon. Not. R. Astron. Soc. **512**(3), 4280–4290 (2022)
126. K. Umetsu, M. Pizzardo, A. Diaferio, M.J. Geller, Cluster Lensing Mass Inversion (CLUMI+): Combining Dynamics and Weak Lensing around Galaxy Clusters (2025). arxiv:2505.04694
127. S. Konrad, C.L. Majer, S. Meyer, E. Sarli, M. Bartelmann, Joint reconstruction of galaxy clusters from gravitational lensing and thermal gas-i. outline of a non-parametric method. Astronomy & Astrophysics 553, 118 (2013)
128. C. Majer, S. Meyer, S. Konrad, E. Sarli, M. Bartelmann, Reconstruction of the mass distribution of galaxy clusters from the inversion of the thermal sunyaev-zel'dovich effect. Mon. Not. R. Astron. Soc. **460**(1), 844–854 (2016)
129. C. Tchernin, C.L. Majer, S. Meyer, E. Sarli, D. Eckert, M. Bartelmann, Reconstructing the projected gravitational potential of abell 1689 from x-ray measurements. Astron. Astrophys. **574**, 122 (2015)
130. C. Tchernin, M. Bartelmann, K. Huber, A. Dekel, G. Hurier, C. Majer, S. Meyer, E. Zinger, D. Eckert, M. Meneghetti, Reconstruction of the two-dimensional gravitational potential of galaxy clusters from x-ray and sunyaev-zel'dovich measurements. Astron. Astrophys. **614**, 38 (2018)
131. S. Bocquet, S. Grandis, L.E. Bleem, M. Klein, J.J. Mohr, T. Schrabback, T.M.C. Abbott, P.A.R. Ade, M. Aguena, A. Alarcon, S. Allam, S.W. Allen, O. Alves, A. Amon, A.J. Anderson, J. Annis, B. Ansarinejad, J.E. Austermann, S. Avila, D. Bacon, M. Bayliss, J.A. Beall, K. Bechtol, M.R. Becker, A.N. Bender, B.A. Benson, G.M. Bernstein, S. Bhargava, F. Bianchini, M. Brodwin, D. Brooks, L. Bryant, A. Campos, R.E.A. Canning, J.E. Carlstrom, A.C. Rosell, M.C. Kind, J. Carretero, F.J. Castander, R. Cawthon, C.L. Chang, C. Chang, P. Chaubal, R. Chen, H.C. Chiang, A. Choi, T.-L. Chou, R. Citron, C.C. Moran, J. Cordero, M. Costanzi, T.M. Crawford, A.T. Crites, L.N. Costa, M.E.S. Pereira, C. Davis, T.M. Davis, J. DeRose, S. Desai, T. Haan, H.T. Diehl, M.A. Dobbs, S. Dodelson, C. Doux, A. Drlica-Wagner, K. Eckert, J. Elvin-Poole, S. Everett, W. Everett, I. Ferrero, A. Ferté, A.M. Flores, J. Frieman, J. Frieman Gallicchio, J. García-Bellido, M. Gatti, E.M. George, G. Giannini, M.D. Gladders, D. Gruen, R.A. Gruendl, N. Gupta, G. Gutierrez, N.W. Halverson, I. Harrison, W.G. Hartley, K. Herner, S.R. Hinton, G.P. Holder, D.L. Hollowood, W.L. Holzapfel, K. Honscheid, J.D. Hrubes, N. Huang, J. Hubmayr, E.M. Huff, D. Huterer, K.D. Irwin, D.J. James, M. Jarvis, G. Khullar, K. Kim, L. Knox, R. Kraft, E. Krause, K. Kuehn, N. Kuropatkin, F. Kéruzoré, O. Lahav, A.T. Lee, P.-F. Leget, D. Li, H. Lin, A. Lowitz, N. MacCrann, G. Mahler, A. Mantz, J.L. Marshall, J. McCullough, M. McDonald, J.J. McMahon, J. Mena-Fernández, F. Menanteau, S.S. Meyer, R. Miquel, J. Montgomery, J. Myles, T. Natoli, A. Navarro-Alsina, J.P. Nibarger, G.I. Noble, V. Novosad, R.L.C. Ogando, Y. Omori, S. Padin, S. Pandey, P. Paschos, S. Patil, A. Pieres, A.A.P. Malagón, A. Porredon, J. Prat, C. Pryke, M. Raveri, C.L. Reichardt, J. Roberson, R.P. Rollins, C. Romero, A. Roodman, J.E. Ruhl, E.S. Rykoff, B.R. Saliwanchik, L. Salvati, C. Sánchez, E. Sanchez, D.S. Cid, A. Saro, K.K. Schaffer, L.F. Secco, I. Sevilla-Noarbe, K. Sharon, E. Sheldon, T. Shin, C. Sievers, G. Smecher, M. Smith, T. Somboonpanyakul, M. Sommer, B. Stalder, A.A. Stark, J. Stephen, V. Strazzullo, E. Suchyta, G. Tarle, C. To, M.A. Troxel, C. Tucker, I. Tutusaus, T.N. Varga, T. Veach, J.D. Vieira, A. Vikhlinin, A. Linden, G. Wang, N. Weaverdyck, J. Weller, N. Whitehorn, W.L.K. Wu, B. Yanny, V. Yefremenko, B. Yin, M. Young, J.A. Zebrowski, Y. Zhang, H. Zohren, J. Zuntz, SPT Clusters with DES and HST Weak Lensing. II. Cosmological Constraints from the Abundance of Massive Halos (2024). arxiv:2401.02075
132. A. Niemiec, M. Jauzac, E. o, M. Limousin, K. Sharon, J.-P. Kneib, P. Natarajan, J. Richard, hybrid-lenstool: a self-consistent algorithm to model galaxy clusters with strong-and weak-lensing simultaneously. Monthly Notices of the Royal Astronomical Society **493**(3), 3331–3340 (2020)
133. A. Niemiec, M. Jauzac, D. Eckert, D. Lagattuta, K. Sharon, A.M. Koekemoer, K. Umetsu, A. Acebron, J.M. Diego, D. Harvey, Beyond the ultradeep frontier fields and legacy observations (buffalo): a high-resolution strong+ weak-lensing view of abell 370. Mon. Not. R. Astron. Soc. **524**(2), 2883–2910 (2023)
134. N.R. Patel, M. Jauzac, A. Niemiec, D. Lagattuta, G. Mahler, B. Beauchesne, A. Edge, H. Ebeling, M. Limousin, The KALEIDOSCOPE survey: a new strong and weak gravitational lensing view of the massive galaxy cluster MACS J1423.8+2404. Monthly Notices of the RAS **533**(4), 4500–4514 (2024). https://doi.org/10.1093/mnras/stae2069. arXiv:2405.04577 [astro-ph.CO]
135. M. Bradac, T. Erben, P. Schneider, H. Hildebrandt, M. Lombardi, M. Schirmer, J.M. Miralles, D. Clowe, S. Schindler, Strong and weak lensing united - ii. the cluster mass distribution of the most x-ray luminous cluster rx j1347.5–1145. Astron. Astrophys. **437**(1), 49–60 (2005). https://doi.org/10.1051/004-6361:20042234
136. J.M. Diego, M. Pascale, B.J. Kavanagh, P. Kelly, L. Dai, B. Frye, T. Broadhurst, Godzilla, a monster lurks in the Sunburst galaxy. Astron. Astrophys. **665**, 134 (2022). https://doi.org/10.1051/0004-6361/202243605. arXiv:2203.08158 [astro-ph.GA]
137. H. Atek, R. Gavazzi, J.R. Weaver, J.M. Diego, T. Schrabback, N.A. Hatch, N. Aghanim, H. Dole, W.G. Hartley, S. Taamoli, G. Congedo, Y. Jimenez-Teja, J.-C. Cuillandre, E. Bañados, S. Belladitta, R.A.A. Bowler, M. Franco, M. Jauzac, G. Mahler, J. Richard, P.-F. Rocci, S. Serjeant, S. Toft, D. Abriola, P. Bergamini, A. Biviano, P. Dimauro, M. Ezziati, J.B. Golden-Marx, C. Grillo, A.C.N. Hughes, Y. Kang, J.-P. Kneib, M. Lombardi, G.A. Mamon, C.J.R. McPartland, M. Meneghetti, H. Miyatake, M. Montes, D.J. Mortlock, P.A. Oesch, N. Okabe, P. Rosati, A.N. Taylor, F. Tarsitano, J. Weller, M. Kluge, R. Laureijs, S. Paltani, T. Saifollahi, M. Schirmer, C. Stone, A. Mora, B. Altieri, A. Amara, S. Andreon, N. Auricchio, M. Baldi, A. Balestra, S. Bardelli, A. Basset, R. Bender, C. Bodendorf, D. Bonino, E. Branchini, M. Brescia, J. Brinchmann, S. Camera, G.P. Candini, V. Capobianco, C. Carbone, V.F. Cardone, J. Carretero, S. Casas, F.J. Castander, M. Castellano, S. Cavuoti, A. Cimatti, C.J. Conselice, L. Conversi, Y. Copin, L. Corcione, F. Courbin, H.M. Courtois, A. Da Silva, H. Degaudenzi, A.M. Di Giorgio, J. Dinis, M. Douspis, F. Dubath, X. Dupac, S. Dusini, A. Ealet, M. Farina, S. Farrens, S. Ferriol, S. Fotopoulou, M. Frailis, E. Franceschi, S. Galeotta, W. Gillard, B. Gillis, C. Giocoli, P. Gómez-Alvarez, A. Grazian, F. Grupp, L. Guzzo, M. Hailey, S.V.H. Haugan, J. Hoar, H. Hoekstra, M.S. Holliman, W. Holmes, I. Hook, F. Hormuth, A. Hornstrup, P. Hudelot, K. Jahnke, M. Jhabvala, E. Keihänen, S. Kermiche, A. Kiessling, T. Kitching, R. Kohley, B. Kubik, K. Kuijken, M. Kümmel, M. Kunz, H.

Kurki-Suonio, O. Lahav, D. Le Mignant, S. Ligori, P.B. Lilje, V. Lindholm, I. Lloro, D. Maino, E. Maiorano, O. Mansutti, O. Marggraf, K. Markovic, N. Martinet, F. Marulli, R. Massey, S. Maurogordato, H.J. McCracken, S. Mei, Y. Mellier, E. Merlin, G. Meylan, M. Moresco, L. Moscardini, R. Nakajima, R.C. Nichol, S.-M. Niemi, C. Padilla, K. Paech, F. Pasian, J.A. Peacock, K. Pedersen, W.J. Percival, V. Pettorino, S. Pires, G. Polenta, M. Poncet, L.A. Popa, L. Pozzetti, F. Raison, A. Renzi, J. Rhodes, G. Riccio, E. Romelli, M. Roncarelli, R. Saglia, D. Sapone, P. Schneider, A. Secroun, G. Seidel, S. Serrano, C. Sirignano, G. Sirri, J. Skottfelt, L. Stanco, P. Tallada-Crespí, H.I. Teplitz, I. Tereno, R. Toledo-Moreo, I. Tutusaus, L. Valenziano, T. Vassallo, G. Verdoes Kleijn, A. Veropalumbo, Y. Wang, E. Zucca, C. Baccigalupi, C. Burigana, G. Castignani, Z. Sakr, V. Scottez, M. Viel, P. Simon, D. Stern, J. Martín-Fleitas, D. Scott, Euclid: Early Release Observations A preview of the Euclid era through a galaxy cluster magnifying lens. arXiv e-prints, 2405–13504 (2024) https://doi.org/10.48550/arXiv.2405.13504. arXiv:2405.13504 [astro-ph.GA]

138. J.M. Diego, G. Congedo, R. Gavazzi, T. Schrabback, H. Atek, B. Jain, J.R. Weaver, Y. Kang, W.G. Hartley, G. Mahler, N. Okabe, J.B. Golden-Marx, M. Meneghetti, J.M. Palencia, M. Kluge, R. Laureijs, T. Saifollahi, M. Schirmer, C. Stone, M. Jauzac, D. Scott, B. Altieri, A. Amara, S. Andreon, N. Auricchio, C. Baccigalupi, M. Baldi, S. Bardelli, P. Battaglia, A. Biviano, E. Branchini, M. Brescia, J. Brinchmann, S. Camera, G. Cañas-Herrera, G.P. Candini, V. Capobianco, C. Carbone, V.F. Cardone, J. Carretero, S. Casas, M. Castellano, G. Castignani, S. Cavuoti, K.C. Chambers, A. Cimatti, C. Colodro-Conde, C.J. Conselice, L. Conversi, Y. Copin, F. Courbin, H.M. Courtois, M. Cropper, J.-C. Cuillandre, A. Da Silva, H. Degaudenzi, G. De Lucia, H. Dole, M. Douspis, F. Dubath, X. Dupac, S. Dusini, S. Escoffier, M. Farina, S. Farrens, F. Faustini, S. Ferriol, F. Finelli, P. Fosalba, N. Fourmanoit, M. Frailis, E. Franceschi, M. Fumana, S. Galeotta, K. George, B. Gillis, C. Giocoli, J. Gracia-Carpio, A. Grazian, F. Grupp, L. Guzzo, S.V.H. Haugan, J. Hoar, W. Holmes, I.M. Hook, F. Hormuth, A. Hornstrup, P. Hudelot, K. Jahnke, M. Jhabvala, B. Joachimi, E. Keihänen, S. Kermiche, M. Kilbinger, B. Kubik, K. Kuijken, M. Kümmel, M. Kunz, H. Kurki-Suonio, A.M.C. Le Brun, D. Le Mignant, S. Ligori, P.B. Lilje, V. Lindholm, I. Lloro, G. Mainetti, D. Maino, E. Maiorano, O. Mansutti, O. Marggraf, M. Martinelli, N. Martinet, F. Marulli, R.J. Massey, E. Medinaceli, S. Mei, M. Melchior, Y. Mellier, E. Merlin, G. Meylan, J.J. Mohr, A. Mora, M. Moresco, L. Moscardini, E. Munari, R. Nakajima, C. Neissner, R.C. Nichol, S.-M. Niemi, C. Padilla, S. Paltani, F. Pasian, K. Pedersen, W.J. Percival, V. Pettorino, S. Pires, G. Polenta, M. Poncet, L.A. Popa, L. Pozzetti, F. Raison, A. Renzi, J. Rhodes, G. Riccio, E. Romelli, M. Roncarelli, R. Saglia, Z. Sakr, D. Sapone, B. Sartoris, P. Schneider, A. Secroun, G. Seidel, M. Seiffert, S. Serrano, P. Simon, C. Sirignano, G. Sirri, L. Stanco, J. Steinwagner, P. Tallada-Crespí, A.N. Taylor, I. Tereno, N. Tessore, S. Toft, R. Toledo-Moreo, F. Torradeflot, I. Tutusaus, L. Valenziano, J. Valiviita, T. Vassallo, G. Verdoes Kleijn, A. Veropalumbo, Y. Wang, J. Weller, G. Zamorani, F.M. Zerbi, E. Zucca, M. Bolzonella, C. Burigana, L. Gabarra, J. Martín-Fleitas, S. Matthew, V. Scottez, M. Sereno, M. Viel, Euclid: Early Release Observations. A combined strong and weak lensing solution for Abell 2390 beyond its virial radius. arXiv e-prints, 2507–08545 (2025) https://doi.org/10.48550/arXiv.2507.08545. arXiv:2507.08545 [astro-ph.CO]
139. J. Liesenborgs, S. De Rijcke, H. Dejonghe, P. Bekaert, Non-parametric inversion of gravitational lensing systems with few images using a multi-objective genetic algorithm. Mon. Not. R. Astron. Soc. **380**(4), 1729–1736 (2007)
140. J. Liesenborgs, S. de Rijcke, H. Dejonghe, GRALE: A genetic algorithm for the non-parametric inversion of strong lensing systems. Astrophysics Source Code Library, record ascl:1011.021 (2010)
141. J. Liesenborgs, L.L. Williams, J. Wagner, S. De Rijcke, Extended lens reconstructions with grale: exploiting time-domain, substructural, and weak lensing information. Mon. Not. R. Astron. Soc. **494**(3), 3253–3274 (2020)
142. D.A. Torres-Ballesteros, L. Castañeda, relensing: Reconstructing the mass profile of galaxy clusters from gravitational lensing. Mon. Not. R. Astron. Soc. **518**(3), 4494–4516 (2023)
143. M. Meneghetti, P. Natarajan, D. Coe, E. Contini, G. De Lucia, C. Giocoli, A. Acebron, S. Borgani, M. Bradac, J.M. Diego, A. Hoag, M. Ishigaki, T.L. Johnson, E. Jullo, R. Kawamata, D. Lam, M. Limousin, J. Liesenborgs, M. Oguri, K. Sebesta, K. Sharon, L.L.R. Williams, A. Zitrin, The Frontier Fields lens modelling comparison project. Monthly Notices of the RAS **472**(3), 3177–3216 (2017). https://doi.org/10.1093/mnras/stx2064. arXiv:1606.04548 [astro-ph.CO]
144. Y.D. Hezaveh, L.P. Levasseur, P.J. Marshall, Fast automated analysis of strong gravitational lenses with convolutional neural networks. Nature **548**(7669), 555–557 (2017). https://doi.org/10.1038/nature23463
145. A. Galan, G. Vernardos, A. Peel, F. Courbin, J.-L. Starck, Using wavelets to capture deviations from smoothness in galaxy-scale strong lenses. Astron. Astrophys. **668**, 155 (2022). https://doi.org/10.1051/0004-6361/202244464
146. L. Biggio, G. Vernardos, A. Galan, A. Peel, F. Courbin, Modeling lens potentials with continuous neural fields in galaxy-scale strong lenses. Astron. Astrophys. **675**, 125 (2023). https://doi.org/10.1051/0004-6361/202245126
147. S. Cha, K. HyeongHan, Z.P. Scofield, H. Joo, M.J. Jee, Precision MARS Mass Reconstruction of A2744: Synergizing the Largest Strong-lensing and Densest Weak-lensing Data Sets from JWST. Astrophys. J. **961**(2), 186 (2024). https://doi.org/10.3847/1538-4357/ad0cbf. arXiv:2308.14805 [astro-ph.GA]
148. S. Cha, B.Y. Cho, H. Joo, W. Lee, K. HyeongHan, Z.P. Scofield, K. Finner, M.J. Jee, A High-Caliber View of the Bullet Cluster Through JWST Strong and Weak Lensing Analyses (2025). arxiv:2503.21870
149. B. Welch, D. Coe, J.M. Diego, A. Zitrin, E. Zackrisson, P. Dimauro, Y. Jiménez-Teja, P. Kelly, G. Mahler, M. Oguri, F.X. Timmes, R. Windhorst, M. Florian, S.E. Mink, R.J. Avila, J. Anderson, L. Bradley, K. Sharon, A. Vikaeus, S. McCandliss, M. Bradač, J. Rigby, B. Frye, S. Toft, V. Strait, M. Trenti, S. Sharma, F. Andrade-Santos, T. Broadhurst, A highly magnified star at redshift 6.2. Nature **603**(7903), 815–818 (2022). https://doi.org/10.1038/s41586-022-04449-y
150. B. Welch, D. Coe, E. Zackrisson, S.E.d. Mink, S. Ravindranath, J. Anderson, G. Brammer, L. Bradley, J. Yoon, P. Kelly, J.M. Diego, R. Windhorst, A. Zitrin, P. Dimauro, Y. Jiménez-Teja, M. Abdurro'uf, A. Nonino, F. Acebron, R.J. Andrade-Santos, M.B. Avila, A. Bayliss, T. Benítez, R. Broadhurst, M. Bhatawdekar, G.B. Bradač, W. Caminha, J. Chen, E. Eldridge, M. Farag, B. Florian, S. Frye, S. Fujimoto, A. Gomez, T.Y.-Y. Henry, T.A. Hsiao, B.L. Hutchison, M. James, I. Joyce, G. Jung, R.L. Khullar, G. Larson, N. Mahler, S. Mandelker, T. McCandliss, R. Morishita, C. Newshore, K. Norman, P.A. O'Connor, M. Oesch, M. Oguri, M. Ouchi, J.R. Postman, R.E. Ryan. Rigby, S. Jr, K. Sharma, V. Sharon, L.-G. Strait, F.X. Strolger, S. Timmes, M. Toft, E. Trenti, A. Vikaeus. Vanzella, Jwst imaging of earendel, the extremely magnified star at redshift z = 6.2. The Astrophysical Journal Letters **940**(1), 1 (2022). https://doi.org/10.3847/2041-8213/ac9d39
151. M. Pascale, L. Dai, B.L. Frye, A.G. Beverage, Is Earendel a Star Cluster?: Metal Poor Globular Cluster Progenitors at $z \sim 6$ (2025). arxiv:2507.05483
152. Z.P. Scofield, M.J. Jee, S. Cha, H. Park, Is Earendel a Star?: Investigating the Sunrise Arc Using JWST Strong and Weak Gravitational Lensing Analyses (2025). arxiv:2504.08879
153. B. Zhao, A. Levis, L. Connor, P.P. Srinivasan, K.L. Bouman, Revealing the 3D Cosmic Web through Gravitationally Constrained Neural Fields (2025). arxiv:2504.15262
154. M. Bradač, T. Treu, D. Applegate, A.H. Gonzalez, D. Clowe, W. Forman, C. Jones, P. Marshall, P. Schneider, D. Zaritsky, Focusing cosmic telescopes: exploring redshift z 5–6 galaxies with the bullet cluster 1e0657- 56. Astrophys. J. **706**(2), 1201 (2009)